\documentclass[prd,aps,superscriptaddress,nofootinbib,amsmath,amssymb,showpacs]{revtex4}
\usepackage{graphicx}
\usepackage{dcolumn}
\usepackage{bm}
\usepackage{multirow}
\usepackage{adjustbox}
\usepackage{textcomp}
\usepackage{mathtext}
\usepackage{hyperref}
\graphicspath{{./eps/}}

\begin{document}

\title{{\Large Hyperon production in quasielastic $\bar{\nu}_{\tau}-$nucleon scattering  }}

\author{A. \surname{Fatima}}
\affiliation{Department of Physics, Aligarh Muslim University, Aligarh-202002, India}
\author{M. Sajjad \surname{Athar}}
\email{sajathar@gmail.com}
\affiliation{Department of Physics, Aligarh Muslim University, Aligarh-202002, India}
\author{S. K. \surname{Singh}}
\affiliation{Department of Physics, Aligarh Muslim University, Aligarh-202002, India}
 
\begin{abstract}
 The theoretical results for the total cross sections and polarization components of the $\tau^{+}$ lepton produced in the 
 charged current induced $|\Delta S| = 1$ quasielastic $\bar\nu_\tau - N$ scattering leading to hyperons~($\Lambda, \Sigma$) 
 have been presented assuming T invariance. The theoretical uncertainties arising due to the use of different vector, axial 
 vector and pseudoscalar form factors as well as the effect of SU(3) symmetry breaking have been studied. We have also 
 presented, for the first time, a comparison of the total cross sections for the production of $e,\mu,\tau$ leptons to 
 facilitate the implications of lepton flavor universality~(LFU) in the $|\Delta S| = 1$ quasielastic reactions induced by 
 the antineutrinos of all flavors {\it i.e.}, $\nu_{l};~l=e,\mu,\tau$.
 \end{abstract}
\pacs{{12.15.-y}, {13.15.+g}, {13.88.+e}} 
 \maketitle
  \section{Introduction}\label{intro}
The ${\tau}$ neutrinos~($\nu_{\tau}$) were experimentally observed for the first time by the DONUT 
collaboration~\cite{Kodama:2000mp, Kodama:2007aa} by observing the $\tau^-$ leptons produced through the charged current 
scattering of $\nu_{\tau}$ on the nucleons in the reaction, $\nu_{\tau} + N \longrightarrow \tau^{-} + X$, where 
$N=n$ or $p$, and $X$ represents hadron(s) in the final state. Since then, a few $\tau^{-}$ production events have been 
observed by the OPERA collaboration at CERN~\cite{Agafonova:2014bcr, Agafonova:2015jxn, Agafonova:2018auq} using accelerator 
neutrinos and by the SuperKamiokande~\cite{Li:2017dbe, Abe:2012jj} as well as the IceCUBE~\cite{Aartsen:2019tjl} 
collaborations using the atmospheric neutrinos where the $\nu_{\tau}$s are assumed to be produced through the $\nu_{\mu} 
\longrightarrow \nu_{\tau}$ oscillation. Since these experiments have observed very few events of the $\tau$ lepton 
production, new experiments for producing larger number of $\tau$ leptons have been proposed by the SHiP~\cite{Yoon:2020yrx, 
Anelli:2015pba, Alekhin:2015byh}, DsTau~\cite{Aoki:2019jry}, DUNE~\cite{Machado:2020yxl, Strait:2016mof, Abi:2018rgm} and 
FASER$\nu$~\cite{FASER:2020gpr} collaborations in which the number of $\tau$ lepton events are expected to reach a few 
hundreds during the running time of 3--5 years of the experiment. The results from these experiments would provide the 
desired data with reasonable statistics on the total and differential scattering cross sections as well as on the 
polarization components of the $\tau$ lepton, which would enable a reliable study of the various aspects and properties of 
$\nu_{\tau}$ and $\tau$ leptons.
 
The $\tau$ lepton production in $\nu_{\tau}-N$ scattering has a threshold of 3.5 GeV in the charged current induced 
quasielastic~(CCQE) reactions {\it i.e.}, $\nu_{\tau} (\bar{\nu}_{\tau}) + N \longrightarrow \tau^{-} (\tau^{+}) + 
N^{\prime}$~($N,N^{\prime} =n$ or $p$). As the energy increases, the production of the $\tau$ leptons is accompanied with the 
inelastic and deep inelastic production of hadrons. In the energy region of $\nu_{\tau}~(\bar{\nu}_{\tau})$ experiments, 
where the energy of the produced $\tau$ lepton is not too large compared to its rest mass $m_{\tau} = 1.776$~GeV, the $\tau$ 
leptons would not be completely longitudinally polarized, and would also have transverse component of the 
polarization~\cite{Fatima:2020pvv, Kuzmin:2003ji, Kuzmin:2004ke, Hagiwara:2003di, Hagiwara:2004gs, Hagiwara:2004xe, 
Sobczyk:2019urm, Valverde:2006yi}. Moreover, any presence of the polarization component perpendicular to the reaction plane 
would provide information about the T noninvariance in $\nu_{\tau}-N$ interactions~\cite{Fatima:2018gjy, Fatima:2018tzs, 
Fatima:2018wsy, Fatima:2021ctt}. The polarization state of the $\tau$ lepton affects the total and differential cross 
sections and is, therefore, an important observable in the study of $\nu_{\tau}-N$ interactions. Thus, it is highly desirable 
that a comprehensive study of the $\tau$ polarization along with the cross sections be made in the quasielastic, inelastic 
and deep inelastic reactions induced by $\nu_{\tau}$ and $\bar{\nu}_{\tau}$ on nucleons in order to understand the 
$\nu_{\tau}-N$ interactions. 
  
In the case of Standard Model~(SM) of particle physics, all the leptons interact with each other through the purely leptonic 
processes and with the quarks through the various semileptonic processes, having the same strength for each leptonic flavor. 
This is called the lepton flavor universality~(LFU) and is an essential feature of the SM. 

The validity of LFU has been experimentally studied in the purely leptonic decays of $\mu$, $\tau$ leptons as well as in the 
various semileptonic decays of mesons and baryons. The LFU seems to work quite well in the case of purely leptonic decays of 
$\mu$ and $\tau$ leptons and $W$ boson~\cite{ATLAS:2021icw}. In the case of semileptonic decays of mesons and baryons like 
$K$, $D$, $D_{s}^{\star}$, $\Lambda$ and $\Lambda_{c}$, etc., involving quark transitions between medium heavy and light 
quarks {\it i.e.}, $s \longrightarrow u l \bar{\nu}_{l}$, $s \longrightarrow d l \bar{l}$, $c \longrightarrow s 
l \bar{\nu}_{l}$, $c \longrightarrow d l \bar{\nu}_{l}$, etc., the available experimental results are in agreement with the 
prediction of SM within statistical uncertainties and no evidence of LFU violation~(LFUV) has been 
reported~\cite{Zhang:2019kky, Yang:2018qdx, Fleischer:2019wlx, Golz:2021yqz, BESIII:2021ynj, BESIII:2018ccy, 
Mezzadri:2018axu}. However, the recent indications of LFUV in semileptonic decays in the heavy quark sector involving 
transitions like $b \longrightarrow s l \bar{l}$ and $b \longrightarrow c l \bar{\nu}_{l}$ at BaBar, LHCb, BESIII and Belle 
collaborations~\cite{Bifani:2018zmi, Albrecht:2021tul, LHCb:2021trn, deSimone:2020kwi, Belle:2019rba, Celani:2021hni} have 
generated great interest in studying the origins of LFUV. Further hints for the LFUV~\cite{Crivellin:2021sff, Bryman:2021teu} 
have been inferred from the recent measurements of the anomalous magnetic moment~($g-2$) of the muons, at the 
Fermilab~\cite{Muong-2:2021ojo}, as well as the anomalies reported in the recent measurements of the Cabibbo 
angle~\cite{Coutinho:2019aiy} and production of lepton pairs through the process of $q\bar{q} \longrightarrow l\bar{l}$~($l=e$ 
and $\mu$) in $pp$ collisions at CERN~\cite{CMS:2021ctt}. In the case of $b \longrightarrow s l \bar{l}$ decays, 
LHCb~\cite{LHCb:2021trn} collaboration has reported the ratio of the branching fractions in $l=\mu$ mode to $l=e$ mode to be 
2.6 $\sigma$ lower than the SM prediction. While in the case of $b \longrightarrow c l \bar{\nu}_{l}$ decays, all the three 
collaborations~\cite{Bifani:2018zmi} have reported the ratio of the branching fraction in $l=\tau$ mode to $l=\mu$ mode, 
which challenge LFU at the level of four standard deviation. This has led to extensive theoretical work in constructing the 
models of new physics~(NP) going beyond the standard model~(BSM) to explain these experimental results on LFUV in the $b$ 
quark sector~\cite{Allwicher:2021ndi, Crivellin:2021bkd, London:2021lfn}. In view of these developments in $b$ quark sector, 
it is natural to apply these theoretical models~\cite{Allwicher:2021ndi, Crivellin:2021bkd, London:2021lfn} in the medium 
heavy quark sector involving $c$ and $s$ quarks decays like $c \longrightarrow s l \bar{\nu}_{l}$, $c \longrightarrow d l 
\bar{\nu}_{l}$, $s \longrightarrow u l \bar{\nu}_{l}$, $s \longrightarrow d l \bar{l}$ and critically study the LFUV effects 
and search for them experimentally~\cite{Zhang:2020dla, Becirevic:2020rzi, Leng:2020fei, Zhang:2019tcs}. While these studies 
are being pursued with some interest, the need for further efforts in this direction has been emphasized 
recently~\cite{BESIII:2021ynj}. 

Notwithstanding the above efforts in the study of LFUV in the semileptonic decays, there have been almost no study of LFUV 
effects in the (anti)neutrino-nucleus scattering. The only exception is comparative study of the quasielastic $\nu_{e} 
(\bar{\nu}_{e})$-nucleon/nucleus and $\nu_{\mu} (\bar{\nu}_{\mu})$-nucleon/nucleus scattering and analyzing the differences 
in the cross sections of these processes arising due to the lepton mass and the radiative corrections and the contributions 
due to the pseudoscalar and second class current form factors~\cite{Akbar:2015yda, Day:2012gb}. A comparison with the 
experimental results would give information about the presence or absence of any LFUV effects, however, the present results 
on the cross sections in the neutrino energy region of a few GeV are not precise enough to conclude about the presence of 
LFUV effects. Such studies of LFUV effects in the (anti)neutrino scattering with $\nu_{l} (\bar{\nu}_{l}); (l=e,\mu,\tau)$ in 
the strange, charm, bottom and top quark sectors have not been done. With the aim of exploring the presence of such effects 
in the strangeness sector, we have studied in some detail the quasielastic scattering process of $\bar{\nu}_{l} + N 
\longrightarrow l^{+} + \Lambda (\Sigma); (l=e, \mu, \tau)$ corresponding to $u\longrightarrow s$ transition. 

In this work, we report on the results of the theoretical calculations within the ambit of the SM with implicit lepton flavor 
universality for the total cross section, differential cross section and the $\tau$ lepton polarization in the reaction
$\bar{\nu}_{\tau} + p \longrightarrow \tau^{+} + \Lambda (\Sigma)$, using various parameterizations for the vector and axial 
vector form factors using SU(3) symmetry. The uncertainties in the numerical values of these observables due to the use of 
different parameterizations of the form factors are discussed. We have also studied the effect of SU(3) symmetry violation 
using the various weak form factors which have been used to study the SU(3) symmetry violating effects in the semileptonic 
hyperon decays of $\Lambda$ and $\Sigma$~\cite{Schlumpf, Faessler:2008ix} and have discussed the uncertainties associated 
with SU(3) violations. Any deviation of the experimental results on the cross sections and $\tau$ polarization to be obtained 
in future will be a signal of LFUV. The results presented here would facilitate the study of LFUV effects in $\bar{\nu}_{l}~
(l=e, \mu, \tau)$ induced processes in the strangeness sector corresponding to $u \longrightarrow s$ transition and 
compliment such LFUV studies in the semileptonic decays of strange particles~\cite{He:2019xxp}. 
    \begin{figure}
 \begin{center}
    \includegraphics[height=3.9cm,width=7.9cm]{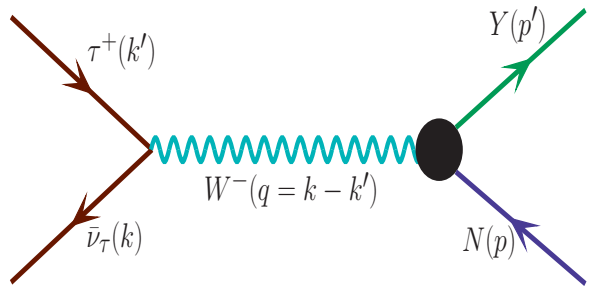}
  \caption{Feynman diagram for the process $\bar{\nu}_\tau (k) + N (p) \rightarrow \tau^+ (k^\prime) + Y (p^\prime)$, where 
  $N=p,n$; $Y=\Lambda,\Sigma^0,\Sigma^-$ and the quantities in the bracket represent four momenta of the corresponding 
  particles.}\label{fyn_hyp}
   \end{center}
 \end{figure}
 
In section~\ref{mat_element}, we present the formalism for calculating the differential and total scattering cross sections 
and using them the polarization observables of the $\tau$ lepton produced in the quasielastic $|\Delta S|=1$ scattering 
processes have been discussed in section-\ref{polarization}. In section~\ref{total_cross_section-1},  we present and discuss 
the numerical results obtained for the differential cross section and polarization observables of the $\tau$ lepton and 
study the dependence of the different parameteric forms of the vector and axial vector form factors as well as the effect of 
SU(3) symmetry breaking on these observables. Similar effects are studied in the case of total cross sections and average 
polarizations in section~\ref{total_cross_section-2}. Further, we have also studied the lepton flavor universality in the 
case of $e-\mu$ and $e-\mu-\tau$ sectors in the total cross sections and the numerical results are presented for the ratios 
$ R_{1} = \frac{\sigma(\bar{\nu}_{\mu} + p \longrightarrow \mu^{+} + \Lambda)}{\sigma(\bar{\nu}_{e} + p \longrightarrow e^{+} 
 + \Lambda)}$ and $R_{2} = \frac{2\sigma(\bar{\nu}_{\tau} + p \longrightarrow \tau^{+} + \Lambda)}{\sigma(\bar{\nu}_{\mu} + p 
 \longrightarrow \mu^{+} + \Lambda) + \sigma(\bar{\nu}_{e} + p \longrightarrow e^{+} + \Lambda)}$, in Section~\ref{LFU_emu}. 
 Section~\ref{summary} summarizes the results and conclude our findings.
   
\section{Formalism}\label{mat_element}
\subsection{Matrix element and weak form factors}
\subsubsection{Matrix element}

The transition matrix element for the quasielastic hyperon production processes depicted in Fig.~\ref{fyn_hyp}, given by
\begin{eqnarray}\label{process1}
 \bar{\nu}_\tau (k) + N (p) &\longrightarrow& \tau^+ (k^\prime) + Y (p^\prime), \qquad N=p,n; \qquad Y =\Lambda,\Sigma^0, 
 \Sigma^-,
\end{eqnarray}
 is written as
 \begin{eqnarray}
 \label{matrixelement}
 {\cal{M}} = \frac{G_F}{\sqrt{2}} \sin \theta_{c}~ l^\mu {{J}}_\mu,
 \end{eqnarray}
 where the quantities in the brackets of Eq.~(\ref{process1}) represent the four momenta of the respective
 particles, $G_F$ is the Fermi coupling constant and $\theta_c~(=13.1^\circ)$ is the Cabibbo mixing angle. The leptonic 
 current $l^\mu$ is given by
 \begin{equation}\label{l}
 l^\mu = \bar{u} (k^\prime) \gamma^\mu (1 + \gamma_5) u (k).
\end{equation}
The hadronic current ${J}_\mu$ is expressed as:
\begin{equation}\label{j}
 {{J}}_\mu =  \bar{u} (p^\prime) {\Gamma_\mu} u (p)
\end{equation}
with
\begin{equation}\label{gamma}
 {\Gamma_\mu} = V_\mu - A_\mu.
\end{equation}
The vector ($V_\mu$) and the axial vector ($A_\mu$) currents are given by~\cite{Fatima:2018tzs, Fatima:2018gjy}:
\begin{eqnarray}\label{vx}
 \langle Y(p^\prime) | V_\mu| N(p) \rangle &=& \bar{u}(p^\prime) \left[ \gamma_\mu f_1^{NY}(Q^2)+i\sigma_{\mu \nu} 
 \frac{q^\nu}{M +M_{Y}} f_2^{NY}(Q^2) + \frac{2 ~q_\mu}{M + M_{Y}} f_3^{NY}(Q^2) \right] u(p),\\
 \label{vy}
  \langle Y(p^\prime) | A_\mu| N(p) \rangle &=& \bar{u} (p^\prime) \left[ \gamma_\mu \gamma_5 g_1^{NY}(Q^2) + 
  i\sigma_{\mu \nu} \frac{q^\nu}{M + M_{Y}} \gamma_5 g_2^{NY}(Q^2) + \frac{2 ~q_\mu} {M + M_{Y}} g_3^{NY}(Q^2) 
  \gamma_5 \right] u(p), 
\end{eqnarray}
where $M$ and $M_{Y}$ are the masses of the initial nucleon and the final hyperon. $q (= k - k^\prime = p^\prime -p)$ is the 
four momentum transfer with $Q^2 = - q^2, Q^2 \ge 0$. $f_1^{NY} (Q^2)$, $f_2^{NY} (Q^2)$ and $f_3^{NY} (Q^2)$ are the $N-Y$ 
transition vector, weak magnetic and induced scalar form factors and $g_1^{NY} (Q^2)$, $g_2^{NY} (Q^2)$ and $g_3^{NY} (Q^2)$ 
are the axial vector, induced tensor~(or weak electric) and induced pseudoscalar form factors, respectively. 

\subsubsection{Weak transition form factors}\label{W_FF}
The weak vector and axial vector form factors are determined using the following assumptions, which are consistent with the 
constraints due to the symmetry properties of the weak hadronic currents~\cite{Marshak, Pais:1971er, LlewellynSmith:1971uhs}:
\begin{itemize}
 \item [(a)] T invariance implies that all the vector~($f_{i}^{NY} (Q^2)$; $i=1-3$) and axial vector~($g_{i}^{NY} (Q^2)$; 
 $i=1-3$) form factors are real.
 
 \item [(b)] The hypothesis that the charged weak vector currents and its conjugate along with the isovector part of the 
 electromagnetic current form an isotriplet implies that the weak vector form factors $f_1^{NY} (Q^2)$ and $f_2^{NY} (Q^2)$ 
 are related to the isovector electromagnetic form factors of the nucleon. The hypothesis ensures conservation of vector 
 current~(CVC) in the weak sector.
 
 \item [(c)] The hypothesis of CVC of the weak vector currents implies that $f_3 (Q^2) = 0$.
 
 \item [(d)] The principle of G-invariance implies the second class current form factors to be zero, {\it i.e.}, $f_3^{NY} 
 (Q^2)= 0$ and $g_2^{NY} (Q^2) = 0$.
 
 \item [(e)] The hypothesis of partially conserved axial vector current~(PCAC) relates the pseudoscalar form 
 factor~($g_{3}^{NY} (Q^2)$) to the axial vector form factor~($g_{1}^{NY} (Q^2)$), through the Goldberger-Treiman~(GT) 
 relation. 
 
  \item [(f)] The assumption of SU(3) symmetry of the weak hadronic currents implies that the vector and axial vector currents 
 transform as an octet under the SU(3) group of transformations.
\end{itemize}
The determination of all the weak form factors is based on the symmetry properties discussed above, and the details are given 
in Ref.~\cite{Fatima:2018tzs, Fatima:2018gjy}. The explicit expressions of the vector and axial vector form factors for the 
different $N-Y$ transitions, assuming the SU(3) symmetry are given in Section~\ref{SU}, while the effects of SU(3) symmetry 
breaking on these form factors are discussed in Section~\ref{SU3_breaking}.
%

\subsubsection{Form factors with SU(3) symmetry}\label{SU} 
The weak vector and the axial vector currents corresponding to the $\Delta S = 1$ currents whose matrix elements are defined 
between the initial~($|N\rangle$) and final~($|Y\rangle$) states in Eq.~(\ref{process1}) are assumed to belong to the octet 
representation of the SU(3). Since $|N\rangle$ and $|Y\rangle$ also belong to the octet representation under SU(3), each of 
these form factors are described in terms of the functions $D(Q^2)$ and $F(Q^2)$ corresponding to the symmetric~(S) and 
antisymmetric~(A) couplings and the SU(3) Clebsch-Gordan coefficients. Explicitly, the form factors can be expressed as~(for 
details, see Ref.~\cite{Fatima:2018tzs}):
\begin{eqnarray}\label{fi}
f_i (Q^2) &=& a F_i^V (Q^2) + b D_i^V  (Q^2)\\
\label{gi}
g_i (Q^2) &=& a F_i^A (Q^2) + b D_i^A  (Q^2) \qquad i =1,2,3
\end{eqnarray}
The Clebsch-Gordan coefficients $a$ and $b$ are calculated for each $N-Y$ transitions and are given in Table~\ref{tabI}. 

\begin{table*}[h]
\begin{tabular}{|c|c|c|}\hline
Transitions ~~ & ~~ $a$ ~~ & ~~ $b$ ~~ \\ \hline
 ~~ $p\rightarrow \Lambda$ ~~ & ~~ $-\sqrt{\frac{3}{2}}$ ~~ & ~~ $-\frac{1}{\sqrt{6}}$ ~~\\
 ~~ $p\rightarrow \Sigma^0$ ~~ & ~~ $-\frac{1}{\sqrt{2}}$ ~~ & ~~ $\frac{1}{\sqrt{2}}$ ~~\\
 ~~ $n\rightarrow \Sigma^-$ ~~ & ~~ $-1$ ~~ & ~~ 1 ~~ \\ \hline
\end{tabular}
\caption{Values of the coefficients $a$ and $b$ given in Eqs.~(\ref{fi})$-$(\ref{gi}).}
\label{tabI}
\end{table*}
From Table~\ref{tabI}, we see that the SU(3) symmetry predicts a relation between the vector and axial vector form factors 
for the transitions $p \longrightarrow \Sigma^{0}$ and $n \longrightarrow \Sigma^{-}$, which implies that 
\begin{equation}\label{sigma_sigma}
 \left[\frac{d\sigma}{dQ^2}\right]_{ p \longrightarrow \Sigma^{0}} = \frac{1}{2} 
  \left[\frac{d\sigma}{dQ^2}\right]_{n \longrightarrow \Sigma^{-}},
\end{equation}
and
\begin{equation}\label{P_sigma}
 \left[P_{L,P}\right]_{p \longrightarrow \Sigma^{0}} = \left[P_{L,P}\right]_{n \longrightarrow \Sigma^{-}}.
\end{equation}

\begin{itemize}
 \item [(i)] {\bf Vector form factors}:\\
 In the case of vector form factors, the functions $D_{i}^{V} (Q^2)$ and $F_{i}^{V} (Q^2)$ are determined in terms of the 
 nucleon electromagnetic form factors $f_{i}^{n,p} (Q^2);~(i=1,2)$, following the same method as discussed 
 above~(Eq.~(\ref{fi})) in the case of electromagnetic interactions, {\it i.e.}
 \begin{eqnarray}\label{fipp}
D_{i}^{V} (Q^2) &=& -\frac{3}{2} f_{i}^{n} (Q^2) \qquad \qquad~~~~~~ \qquad i=1,2 \\
\label{finn}
F_i^{V} (Q^2) &=& f_{i}^{p} (Q^2) + \frac{1}{2} f_i^n (Q^2)  \qquad ~~\qquad i =1,2
 \end{eqnarray}
Using the expressions of $D_{i}^{V} (Q^2)$ and $F_{i}^{V} (Q^2)$ obtained above and the values of the coefficients $a$ and 
$b$ from Table~\ref{tabI}, the vector form factors $f_{1,2}^{NY} (Q^2)$ are expressed in terms of the nucleon electromagnetic 
form factors as:
\begin{eqnarray}\label{f21}
 f_{1,2}^{p \Lambda}(Q^2)&=& -\sqrt{\frac{3}{2}}~f_{1,2}^p(Q^2), \\
 f_{1,2}^{n \Sigma^-}(Q^2)&=& -\left[f_{1,2}^p(Q^2) + 2 f_{1,2}^n(Q^2) \right],
 \end{eqnarray}
 \begin{eqnarray}
 \label{f2n}
  f_{1,2}^{p \Sigma^0}(Q^2)&=& -\frac{1}{\sqrt2}\left[f_{1,2}^p(Q^2) + 2 f_{1,2}^n(Q^2) \right].
\end{eqnarray}

The electromagnetic nucleon form factors, in turn, are expressed in the terms of the Sachs' electric~($G_{E} (Q^2)$) and 
magnetic~($G_{M} (Q^2)$) form factors of the nucleons, for which various parameterizations are available in the 
literature~\cite{Bradford:2006yz, Bosted:1994tm, Budd:2004bp,Alberico:2008sz, Kelly:2004hm, Galster:1971kv, Platchkov:1989ch, 
Punjabi:2015bba}. For the numerical calculations, we have used the parameterization given by Bradford {\it et 
al.}~(BBBA05)~\cite{Bradford:2006yz} unless stated otherwise.
 
 \item [(ii)] {\bf Axial vector form factors}:\\
We express $g_{1}^{NY}(Q^2)$ in terms of $g_{1} (Q^2) $ and $x_{1} (Q^2)$, which are 
defined as
\begin{eqnarray}\label{gnp}
  g_1 (Q^2) &=& F_{1}^A (Q^2) + D_{1}^A (Q^2), \\
  x_{1} (Q^2) &=& \frac{F_{1}^A (Q^2)}{F_{1}^A (Q^2) + D_{1}^A (Q^2)}; 
 \end{eqnarray}
 where $F_{1}^{A} (Q^2)$ and $D_{1}^{A} (Q^2)$ are the antisymmetric and symmetric couplings of the two octets, determined 
 from the semileptonic decays of hyperons at very low $Q^2$ $\approx 0$. It may be pointed out that there is no information 
 available in the literature, for the $Q^2$ dependence of these parameters. Therefore, phenomenologically same $Q^2$ 
 dependence for $F_{1}^{A} $ and $D_{1}^{A}$, that is the dipole form is assumed, such that the parameter $x_{1} (Q^2)$ 
 becomes a constant, {\it i.e.,} $x_{1} (Q^2) \approx x_{1} (0) = 0.364$.

The explicit expressions of $g_1^{NY} (Q^2)$ for $p \longrightarrow \Lambda$, $p \longrightarrow \Sigma^0$ and 
$n \longrightarrow \Sigma^{-}$ are given as
   \begin{eqnarray}\label{gplam}
 g_{1}^{p \Lambda}(Q^2)&=& -\frac{1}{\sqrt{6}} (1+2x_{1}) g_{1} (Q^2), \\
 \label{gnsig}
 g_{1}^{n \Sigma^-}(Q^2)&=&   (1-2x_{1}) g_{1}(Q^2),\\
 \label{gpsig}
 g_{1}^{p \Sigma^0}(Q^2)&=& \frac{1}{\sqrt2}(1-2x_{1})g_{1}(Q^2),
\end{eqnarray}
where 
\begin{equation}\label{ga}
  g_1 (Q^2) = \frac{g_A (0)}{\left( 1 + \frac{Q^2}{M_A^2} \right)^2},
  \end{equation}
  with $g_A(0) = 1.267$~\cite{Cabibbo:2003cu} and $M_A = 1.026$ GeV~\cite{Bernard:2001rs}. 
  
\item [(iii)] {\bf Pseudoscalar form factor}:\\
The contribution of $g_3 ^{NY}(Q^2)$ in $\nu_{\tau}~(\bar{\nu}_{\tau}) - N$ scattering is significant due to the high value 
of $m_\tau$. In literature, there exists two parameterizations, given by Marshak {\it et al.}~\cite{Marshak} and by 
Nambu~\cite{Nambu:1960xd}, for the pseudoscalar form factor in the $\Delta S =1$ channel. In order to study the effect of 
the pseudoscalar form factor on the cross section and polarization observables, we have used both the parameterizations. The 
expression of the pseudoscalar form factor parameterized by Nambu~\cite{Nambu:1960xd} is given as:
\begin{equation}\label{g3}
  g_3^{NY}(Q^2)=\frac{(M+M_{Y})^2 }{2(m_{K}^2+Q^2)} g_1^{NY}(Q^2),
\end{equation}
where $m_{K}$ is the kaon mass.
 
In the parameterization of Marshak {\it et al.}~\cite{Marshak}, the expression for the pseudoscalar form factor is given as:
\begin{equation}\label{g3_Marshak}
g_3^{NY}(Q^2)=\frac{(M+M_{Y})^2 }{2Q^2} \frac{g_{1}^{NY} (Q^2) (m_{K}^{2} + Q^2) - m_{K}^{2} g_{1}^{NY} (0)}{m_{K}^{2} + Q^2}.
\end{equation}
\end{itemize}

\subsubsection{Form factors with SU(3) symmetry breaking effects}\label{SU3_breaking}
In the literature, SU(3) symmetry breaking effects have been studied by various groups~\cite{Wang:2019alu, Shanahan:2015dka, 
Yang:2015era, Becirevic:2004bb, Lacour:2007wm, Pham:2012db, Pham:2014lla, Ledwig:2014rfa, Chang:2014iba} 
especially in the case of semileptonic decays of hyperons. In this work, we have studied the effect of SU(3) symmetry 
breaking parameterized in the two models by Faessler {\it et al.}~\cite{Faessler:2008ix} and Schlumpf~\cite{Schlumpf}:
 \begin{itemize}
  \item [(A)] {\bf Faessler et al.}~\cite{Faessler:2008ix}:\\
 The main features of the model may be summarized as:
 \begin{itemize}
  \item [i)] At the leading order, there is no symmetry breaking effect for the vector form factor $f_{1}^{NY}(Q^{2})$ 
  because of the  Ademollo-Gatto theorem~\cite{Ademollo:1964sr}.
  
  \item [ii)] In the presence of SU(3) symmetry breaking, the value of $f_{2}^{NY} (Q^{2})$ is modified from its SU(3) 
  symmetric value $f_{2}^{NY} (Q^2)$ to ${\cal F}_{2}^{NY}(Q^2)$, as
  \begin{eqnarray}\label{SU3}
   f_{2}^{p\Lambda} (Q^2) &\longrightarrow& {\cal F}_{2}^{p\Lambda} (Q^2) = f_{2}^{p\Lambda} (Q^2) -{\frac{1}{3\sqrt{6}}} 
   \left[H_{1}^{V} (Q^2) -2H_{2}^{V} (Q^2) -3H_{3}^{V} (Q^2) -6H_{4}^{V} (Q^2)  \right],\\
   f_{2}^{n\Sigma^{-}} (Q^2) &\longrightarrow& {\cal F}_{2}^{n\Sigma^{-}}(Q^2) = f_{2}^{n\Sigma^{-}} (Q^2) -\frac{1}{3} 
   \left[H_{1}^{V} (Q^2) + H_{3}^{V} (Q^2) \right], \\
   f_{2}^{p\Sigma^{0}} (Q^2) &\longrightarrow& {\cal F}_{2}^{p\Sigma^{0}}(Q^2) = f_{2}^{p\Sigma^{0}} (Q^2) - 
   \frac{1}{3\sqrt{2}} \left[H_{1}^{V} (Q^2) + H_{3}^{V} (Q^2) \right],
    \end{eqnarray}
where $f_{2}^{NY} (Q^2)$ for the different $N-Y$ transitions are given in Eqs.(\ref{f21})--(\ref{f2n}) and $H_{i}^{V} (Q^2); 
i=1-4$ are the SU(3) symmetry breaking terms. Since the symmetry breaking effects, in this model, are studied for the 
semileptonic decays of hyperons at very low $Q^2$, {\it i.e.}, $Q^2 \simeq 0$, therefore, no information about the $Q^2$ 
dependence of $H_{i}^{V} (Q^2)$ is available in the literature. For simplicity, a dipole parameterization is assumed
\begin{equation}
 H_{i}^{V} (Q^2) = \frac{H_{i}^{V} (0)}{\left(1+\frac{Q^2}{M_{V}^2}\right)^2}, \qquad \quad i=1-4
\end{equation}
where $M_{V} =0.84$~GeV is the vector dipole mass, and the values of the couplings $H_{i}^{V} (0)$ are given in 
Ref.~\cite{Faessler:2008ix}, and are here quoted as:
\begin{eqnarray}
 H_{1}^{V}(0) ~=~ -0.246, \qquad \quad
 H_{2}^{V}(0) ~=~ 0.096, \qquad \quad H_{3}^{V}(0) ~=~ 0.021, \qquad \quad H_{4}^{V}(0) ~=~ 0.030 \nonumber 
\end{eqnarray}
Similarly, the axial vector form factor $g_{1}^{NY}(Q^2)$, in the presence of SU(3) symmetry breaking, is modified to 
${\cal G}_{1}^{NY} (Q^2)$ as
  \begin{eqnarray}\label{SU3_g1}
   g_{1}^{p\Lambda} (Q^2) &\longrightarrow& {\cal G}_{1}^{p\Lambda}(Q^2) = g_{1}^{p\Lambda} (Q^2) -{\frac{1}{3\sqrt{6}}} 
   \left[H_{1}^{A} (Q^2) -2H_{2}^{A} (Q^2) -3H_{3}^{A} (Q^2) -6H_{4}^{A} (Q^2) \right],\\
   g_{1}^{n\Sigma^{-}} (Q^2) &\longrightarrow& {\cal G}_{1}^{n\Sigma^{-}}(Q^2) = g_{1}^{n\Sigma^{-}} (Q^2) -\frac{1}{3}
   \left[H_{1}^{A} (Q^2) + H_{3}^{A} (Q^2) \right], \\
   g_{1}^{p\Sigma^{0}} (Q^2) &\longrightarrow& {\cal G}_{1}^{p\Sigma^{0}}(Q^2) = g_{1}^{p\Sigma^{0}} (Q^2) + \frac{1}
   {3\sqrt{2}} \left[H_{1}^{A} (Q^2) + H_{3}^{A} (Q^2) \right],
    \end{eqnarray}
with $g_{1}^{NY} (Q^2)$ defined in Eqs.(\ref{gplam})--(\ref{gpsig}) and a dipole parameterization is assumed for $H_{i}^{A} 
(Q^2)$ as
\begin{equation}
 H_{i}^{A} (Q^2) = \frac{H_{i}^{A} (0)}{\left(1+\frac{Q^2}{M_{A}^2}\right)^2}; \qquad \quad i=1-4
\end{equation}
where the couplings $H_{i}^{A} (0)$ are~\cite{Faessler:2008ix}
  \begin{eqnarray}
 H_{1}^{A} (0) ~=~ -0.050, \qquad \quad
 H_{2}^{A} (0) ~=~ 0.011, \qquad \quad H_{3}^{A}(0) ~=~ -0.006, \qquad \quad H_{4}^{A}(0) ~=~ 0.037 .\nonumber 
\end{eqnarray}

  \item [iii)] Since the pseudoscalar form factor is parameterized in terms of the axial vector form 
  factor~(Eqs.~(\ref{g3}) and (\ref{g3_Marshak})), therefore, it receives  SU(3) symmetry breaking effect {\it via.}, 
  $g_{1}^{NY} (Q^2)$.
 \end{itemize}

 \item [(B)] {\bf Schlumpf}~\cite{Schlumpf}:\\
 Schlumpf~\cite{Schlumpf} has studied SU(3) symmetry breaking in the hadronic current containing vector~($f_1$) and axial 
 vector~($g_1$) form factors using relativistic quark model, and this symmetry breaking in the model originates from the mass 
 difference between $m_s$ and $m_{u/d}$ quarks. The modified $f_1$ and $g_1$ form factors are given by
 \begin{eqnarray}
  f_1(Q^2) &\rightarrow& f_1^\prime(Q^2)=\alpha f_1(Q^2)\nonumber\\
 g_1(Q^2) &\rightarrow& g_1^\prime(Q^2)=\beta g_1(Q^2), \text{where}
 \end{eqnarray}
 $\alpha$=0.976, 0.975 and 0.975;~~$\beta$=1.072, 1.051 and 1.056, respectively for $p \longrightarrow \Lambda$, $p 
 \longrightarrow \Sigma^0$ and $n \longrightarrow \Sigma^-$ transitions. Since the induced pseudoscalar form factor $g_3$ is 
 related to the axial vector form factor $g_1$, therefore, this modification is also applicable to the terms containing 
 $g_3$ through Eqs.~(\ref{g3}) and (\ref{g3_Marshak}).
 \end{itemize}
 
\section{Cross section and polarization observables of the final lepton}\label{polarization}
\subsection{Cross section}
The general expression of the differential cross section for the processes given in Eq.~(\ref{process1}), in the laboratory 
frame, 
is given by
 \begin{eqnarray}
 \label{crosv.eq}
 d\sigma&=&\frac{1}{(2\pi)^2}\frac{1}{4 M E_{\bar{\nu}_{\tau}}}\delta^4(k+p-k^\prime-p^\prime) 
 \frac{d^3k^\prime}{2E_{k^\prime}} \frac{d^3p^\prime}{2E_{p^\prime}} \overline{\sum} \sum |{\cal{M}}|^2.
 \end{eqnarray}

 Using Eqs.~(\ref{matrixelement})--(\ref{j}), the transition matrix element squared is obtained as:
\begin{equation}\label{matrix}
  \overline{\sum} \sum |{\cal{M}}|^2 = \frac{G_F^2 \sin^2 \theta_c}{2} \cal{J}^{\mu \nu} \cal{L}_{\mu \nu},
\end{equation} 
where the hadronic~($\cal{J}_{\mu \nu}$) and the leptonic~($\cal{L}_{\mu \nu}$) tensors are obtained using Eqs.~(\ref{l}) 
and (\ref{j}) as
\begin{eqnarray}\label{J}
\cal{J}_{\mu \nu} &=& \overline{\sum} \sum J_\mu J_\nu^\dagger, \qquad \quad {\cal{L}}^{\mu \nu} ~=~ \overline{\sum} \sum 
l_\mu l_\nu^\dagger.
\end{eqnarray} 
 
 Following the above definitions, the differential scattering cross section $d\sigma/dQ^2$ for the processes given in 
 Eq.~(\ref{process1}) is written as
\begin{equation}\label{dsig}
 \frac{d\sigma}{dQ^2}=\frac{G_F^2 \sin^2\theta_c}{8 \pi {M}^2 {E^2_{\bar{\nu}_{\tau}}}} N(Q^2),
\end{equation}
where $N(Q^2) = \cal{J}^{\mu \nu} \cal{L}_{\mu \nu}$ is obtained from the expression given
 in Appendix-A of Ref.~\cite{Fatima:2018tzs} with the substitution of $M^{\prime} =M_{Y}$ and $m_{\mu}=m_{\tau}$.

 \subsection{Polarization of the final lepton}
Using the covariant density matrix formalism, the polarization 4-vector~($\zeta^\tau$) of the $\tau$ lepton produced in the 
final state in reactions given in Eq.~(\ref{process1}) is written as~\cite{Bilenky}
\begin{equation}\label{polarl}
\zeta^{\tau}=\frac{\mathrm{Tr}[\gamma^{\tau}\gamma_{5}~\rho_{f}(k^\prime)]}
{\mathrm{Tr}[\rho_{f}(k^\prime)]},
\end{equation}
and the spin density matrix for the final lepton $\rho_f(k^\prime)$ is given by 
\begin{equation}\label{polar1l}
 \rho_{f}(k^\prime)= {\cal J}^{\alpha \beta}  \text{ Tr}[\Lambda(k') \gamma_\alpha (1 \pm \gamma_5) \Lambda(k) 
 \tilde\gamma_ {\beta} (1 \pm \tilde\gamma_5)\Lambda(k')], 
\end{equation} 
with $\tilde{\gamma}_{\alpha} =\gamma^0 \gamma^{\dagger}_{\alpha} \gamma^0$ and $\tilde{\gamma}_{5} =\gamma^0 
\gamma^{\dagger}_{5} \gamma^0$.
 \begin{figure}
 \begin{center}  
  \includegraphics[height=8cm,width=13cm]{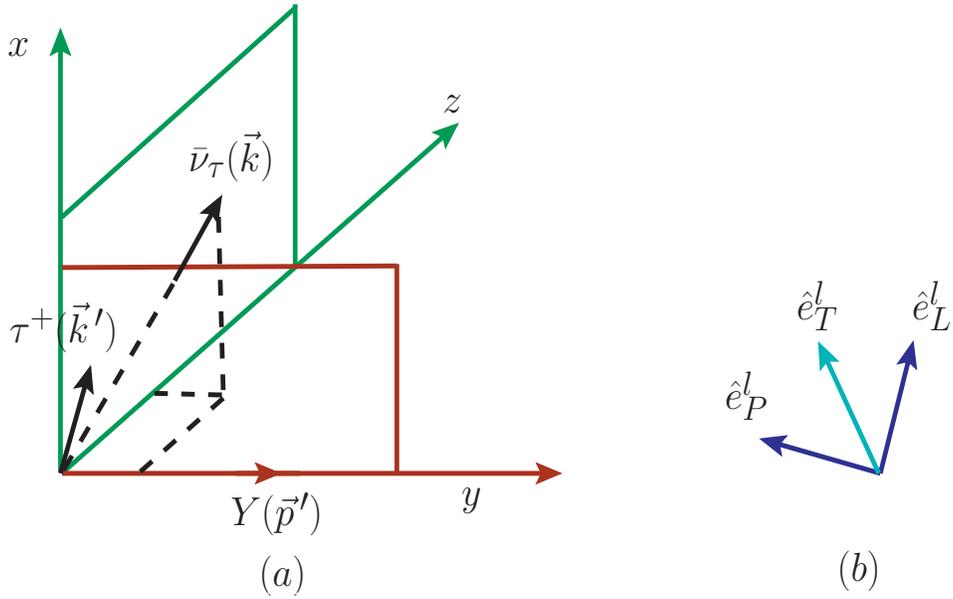}
  \caption{(a) Momentum and polarization directions of the final lepton produced in the reaction $\bar{\nu}_{\tau} (k) + N (p)
  \longrightarrow \tau^{+} (k^{\prime}) + Y (p^{\prime})$. (b)~$\hat{e}_{L}^{l}$, $\hat{e}_{P}^{l}$ and $\hat{e}_{T}^{l}$ 
  represent the orthogonal unit vectors corresponding to the longitudinal, perpendicular and transverse directions with 
  respect to the momentum of the final lepton.}\label{T invariance}
 \end{center}
 \end{figure}

Using the following relations:
\begin{equation}\label{polar3}
\Lambda(k')\gamma^{\tau}\gamma_{5}\Lambda(k')=2m_{\tau} \left(g^{\tau\sigma}-\frac{k'^{\tau}k'^{\sigma}}{m_{\tau}^{2}}
\right)\Lambda(k')\gamma_{\sigma}\gamma_{5}
\end{equation}
and
\begin{equation}\label{polar31}
 \Lambda(k^\prime)\Lambda(k^\prime) = 2m_{\tau} \Lambda(k^\prime),
\end{equation}
$\zeta^\tau$ defined in Eq.~(\ref{polarl}) may also be rewritten as
\begin{equation}\label{polar4l}
\zeta^{\tau}=\left( g^{\tau\sigma}-\frac{k'^{\tau}k'^{\sigma}}{m_\tau^2}\right)
\frac{  {\cal J}^{\alpha \beta}  \mathrm{Tr}
\left[\gamma_{\sigma}\gamma_{5}\Lambda(k') \gamma_\alpha (1 + \gamma_5) \Lambda(k) \tilde\gamma_ {\beta} (1 +
\tilde\gamma_5) \right]}
{ {\cal J}^{\alpha \beta} \mathrm{Tr}\left[\Lambda(k') \gamma_\alpha (1 + \gamma_5) \Lambda(k) \tilde\gamma_ {\beta} 
(1 + \tilde\gamma_5) \right]},
\end{equation}
where $m_\tau$ is the mass of the $\tau$ lepton. In Eq.~(\ref{polar4l}), the denominator is directly related to the 
differential cross section given in Eq.~(\ref{dsig}).

With ${\cal J}^{\alpha \beta}$ and ${\cal L}_{\alpha \beta}$ given in Eq.~(\ref{J}), an expression for $\zeta^\tau$ is 
obtained. In the laboratory frame where the initial nucleon is at rest, the polarization vector $\vec{\zeta}$, assuming T 
invariance, is calculated to be a function of 3-momenta of incoming antineutrino $({\vec{k}})$ and outgoing lepton 
$({\vec{k}}\,^{\prime})$, and is given as  
\begin{equation}\label{3poll}
 \vec{\zeta} =\left[{A^l(Q^2)\vec{ k}} + B^l(Q^2){\vec{k}}\,^{\prime} \right], 
\end{equation}
where the expressions of $A^l(Q^2)$ and $B^l(Q^2)$ are obtained from the expression given in Appendix-B of 
Ref.~\cite{Fatima:2018tzs} with the substitution $M^{\prime} =M_{Y}$ and $m_{\mu}=m_{\tau}$.

One may expand the polarization vector $\vec{\zeta}$ along the orthogonal directions, ${\hat{e}}_L^l$, ${\hat{e}}_P^l$ and 
${\hat{e}_T^l}$ in the reaction plane corresponding to the longitudinal, perpendicular and transverse directions of the final 
lepton~($l$), as depicted in Fig.~\ref{T invariance} and defined as
\begin{equation}\label{vectorsl}
\hat{e}_{L}^l=\frac{\vec{ k}^{\, \prime}}{|\vec{ k}^{\,\prime}|},\qquad
\hat{e}_{P}^l=\hat{e}_{L}^l \times \hat{e}_T^l,\qquad   {\rm where}~~~~~ 
\hat{e}_T^l=\frac{\vec{ k}\times \vec{ k}^{\,\prime}}{|\vec{ k}\times \vec{ k}^{\,\prime}|}.
\end{equation}
We then write $\vec{\zeta}$ as:
 \begin{equation}\label{polarLabl}
\vec{\zeta}=\zeta_{L} \hat{e}_{L}^l+\zeta_{P} \hat{e}_{P}^l + \zeta_{T} \hat{e}_{T}^l,
\end{equation}
such that the longitudinal and  perpendicular components of the $\vec{\zeta}$ in the laboratory frame are 
given by
\begin{equation}\label{PLl}
 \zeta_L(Q^2)=\vec{\zeta} \cdot \hat{e}_L^l,\qquad \zeta_P(Q^2)= \vec{\zeta} \cdot \hat{e}_P^l.
\end{equation}
From Eq.~(\ref{PLl}), the longitudinal $P_L (Q^2)$ and perpendicular $P_P(Q^2)$  
components of the polarization vector defined in the rest frame of the outgoing lepton are given by 
\begin{equation}\label{PL1l}
 P_L(Q^2)=\frac{m_\tau}{E_{k^\prime}} \zeta_L(Q^2), \qquad P_P(Q^2)=\zeta_P(Q^2),
\end{equation}
where $\frac{m_\tau}{E_{k^\prime}}$ is the Lorentz boost factor along ${\vec k}\, ^\prime$. Using Eqs.~(\ref{3poll}), 
(\ref{vectorsl}) and (\ref{PLl}) in Eq. (\ref{PL1l}), the longitudinal $P_L(Q^2)$ and perpendicular 
$P_P (Q^2)$ components are calculated to be
\begin{eqnarray}
  P_L (Q^2) &=& \frac{m_\tau}{E_{k^{\prime}}} \frac{A^l(Q^2) \vec{k}.\vec{k}^{\,\prime} + B^l (Q^2) 
  |\vec{k}^{\,\prime}|^2}{N(Q^2)~|\vec{k}^{\,\prime}|},\label{Pll} \\
 P_P (Q^2) &=& \frac{A^l(Q^2) [|\vec{k}|^2 |\vec{k}^{\,\prime}|^2 - (\vec{k}.\vec{k}^{\,\prime})^2]}{N(Q^2)~
 |\vec{k}^{\,\prime}| ~ |\vec{k}\times \vec{k}^{\,\prime}|},\label{Ppl}
\end{eqnarray}
where $N(Q^2) = \cal{J}^{\mu \nu} \cal{L}_{\mu \nu}$ is obtained from the expression given
 in Appendix-A of Ref.~\cite{Fatima:2018tzs}.

\section{Results and discussion}
In this section, we present and discuss the results of the differential~(Section~\ref{total_cross_section-1}) and 
total~(Section~\ref{total_cross_section-2}) scattering cross sections as well as the polarization observables of the final 
$\tau$ lepton produced in the $|\Delta S|=1$ quasielastic scattering of $\bar{\nu}_{\tau}$ from nucleons. We also present a 
comparison of the total cross section for the production of $e$, $\mu$ and $\tau$ leptons in the quasielastic scattering of 
$\bar{\nu}_{e}$, $\bar{\nu}_{\mu}$ and $\bar{\nu}_{\tau}$ to demonstrate the implications of LFU in these 
processes~(Section~\ref{LFU_emu}).

\subsection{Differential scattering cross section and polarization observables}
\label{total_cross_section-1}
\begin{figure}
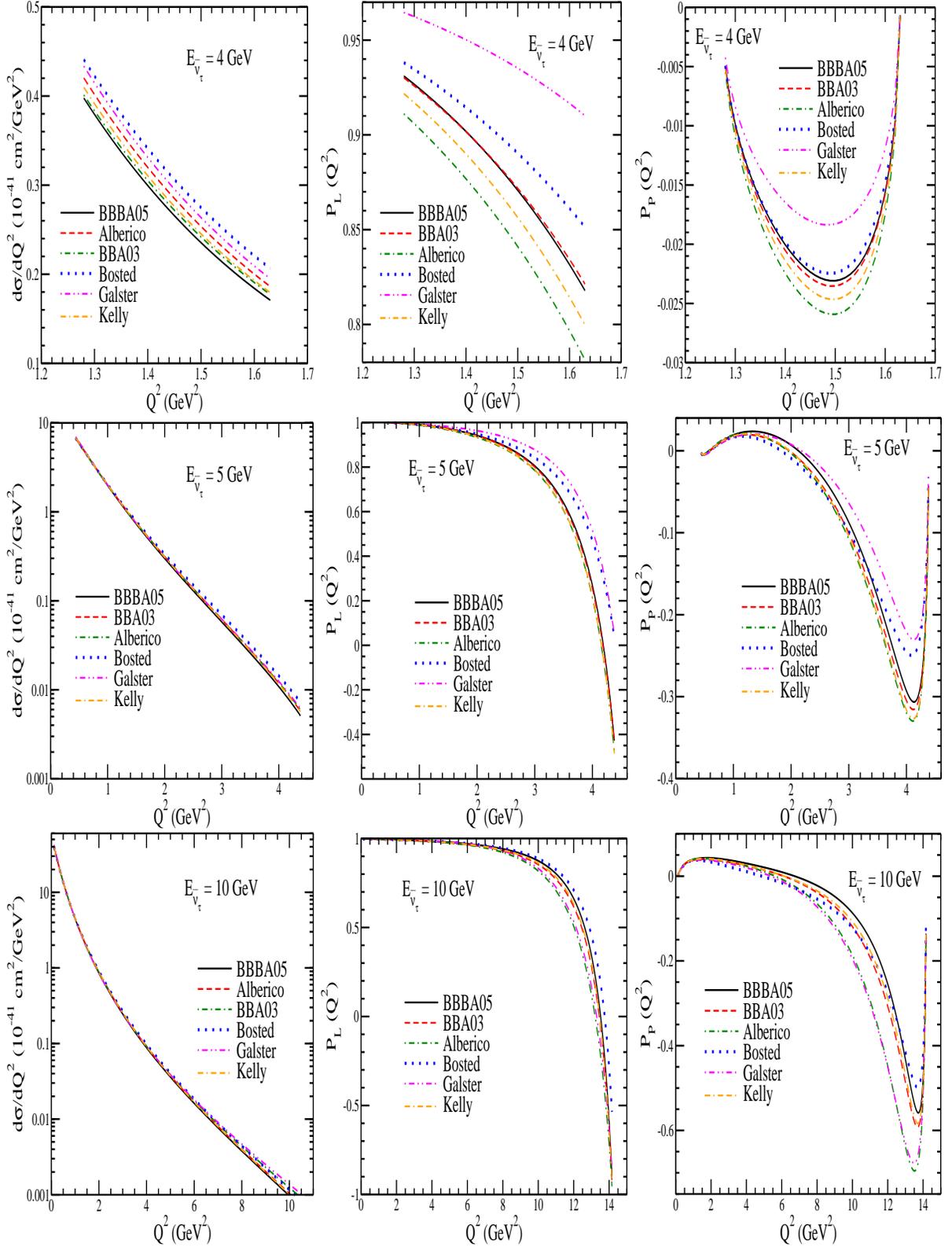

\centering
\includegraphics[height=7cm,width=5.2cm]{dsig_dq2_FF_variation_enu_4GeV_lambda_polarized.eps}
\includegraphics[height=7cm,width=5.2cm]{Pl_q2_FF_variation_enu_4GeV_lambda.eps}
\includegraphics[height=7cm,width=5.2cm]{Pp_q2_FF_variation_enu_4GeV_lambda.eps}
\includegraphics[height=7cm,width=5.2cm]{dsig_dq2_FF_variation_enu_5GeV_lambda_polarized.eps}
\includegraphics[height=7cm,width=5.2cm]{Pl_q2_FF_variation_enu_5GeV_lambda.eps}
\includegraphics[height=7cm,width=5.2cm]{Pp_q2_FF_variation_enu_5GeV_lambda.eps}
\includegraphics[height=7cm,width=5.2cm]{dsig_dq2_FF_variation_enu_10GeV_lambda_polarized.eps} 
\includegraphics[height=7cm,width=5.2cm]{Pl_q2_FF_variation_enu_10GeV_lambda.eps}
\includegraphics[height=7cm,width=5.2cm]{Pp_q2_FF_variation_enu_10GeV_lambda.eps}
\caption{$\frac{d\sigma}{dQ^2}$~(left panel), $P_L (Q^2)$~(middle panel) and $P_P (Q^2)$~(right panel) versus $Q^2$ for the 
process $\bar{\nu}_{\tau} + p \longrightarrow \tau^+ + \Lambda$ at $E_{\bar{\nu}_{\tau}}$ = 4 GeV~(upper panel), 5 GeV~(middle 
panel) and 10 GeV~(lower panel). The calculations have been performed using the SU(3) symmetry with the axial dipole mass 
$M_{A} = 1.026$~GeV and for the different parameterizations of the nucleon vector form factors {\it viz.}, 
BBBA05~\cite{Bradford:2006yz}~(solid line), BBA03~\cite{Budd:2004bp}~(dashed line), 
Alberico {\it et al.}~\cite{Alberico:2008sz}~(dashed-dotted line), Bosted~\cite{Bosted:1994tm}~(dotted line), Galster {\it et 
al.}~\cite{Galster:1971kv}~(double-dotted-dashed line) and Kelly~\cite{Kelly:2004hm}~(double-dashed-dotted line).}
\label{dsigma_pol_VFF_lambda}
\end{figure}
\begin{figure}
\centering
\includegraphics[height=7cm,width=5.2cm]{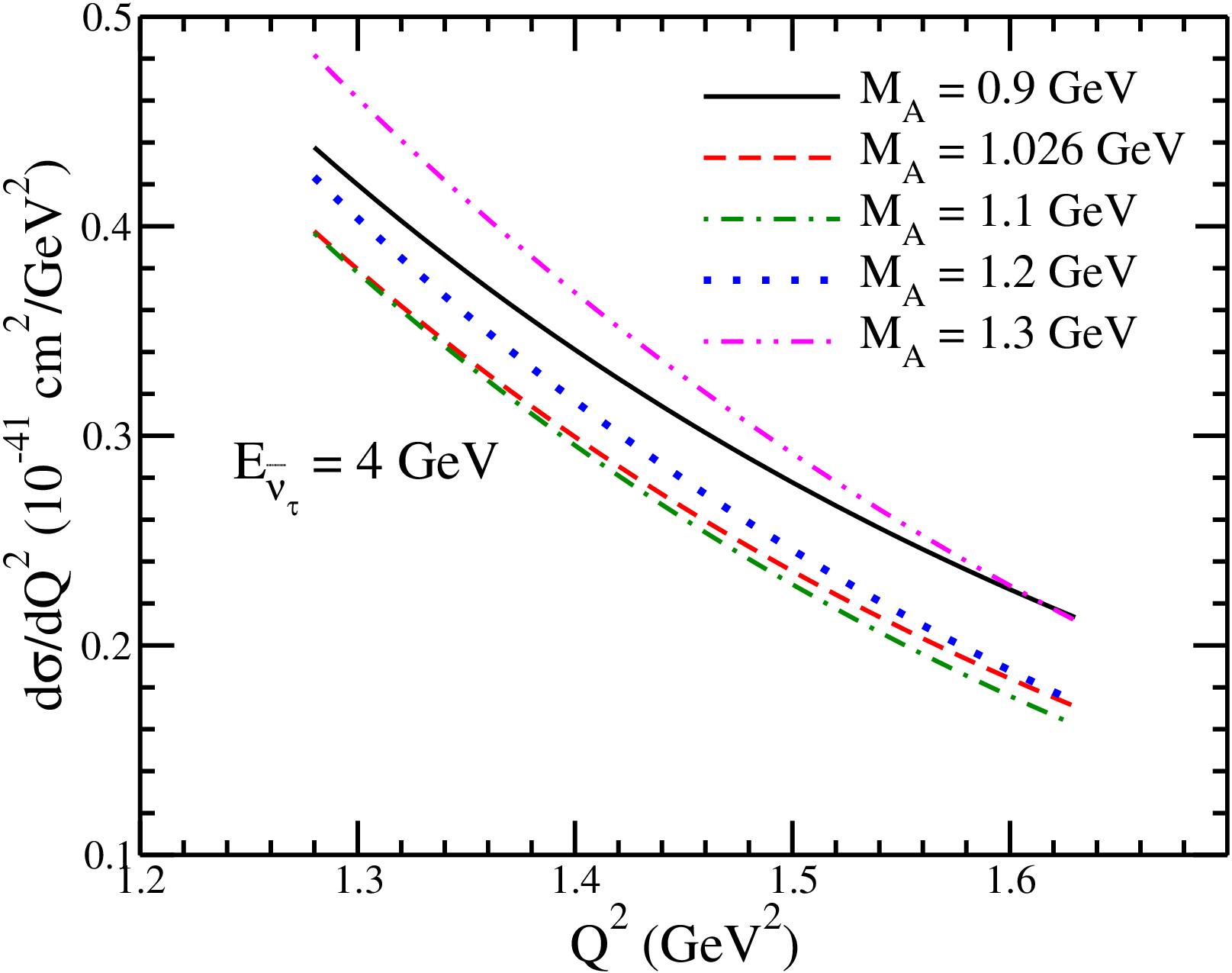}
\includegraphics[height=7cm,width=5.2cm]{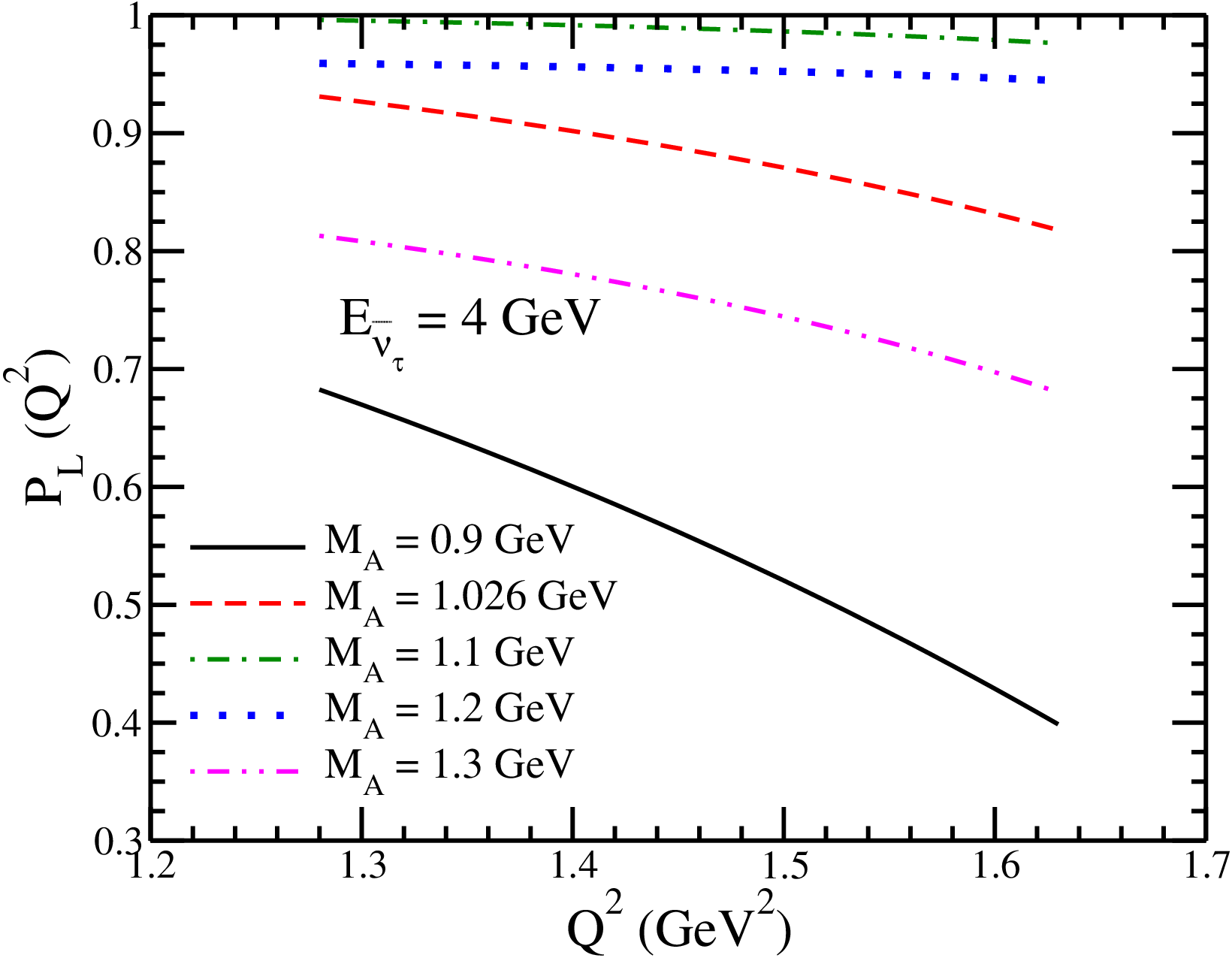}
\includegraphics[height=7cm,width=5.2cm]{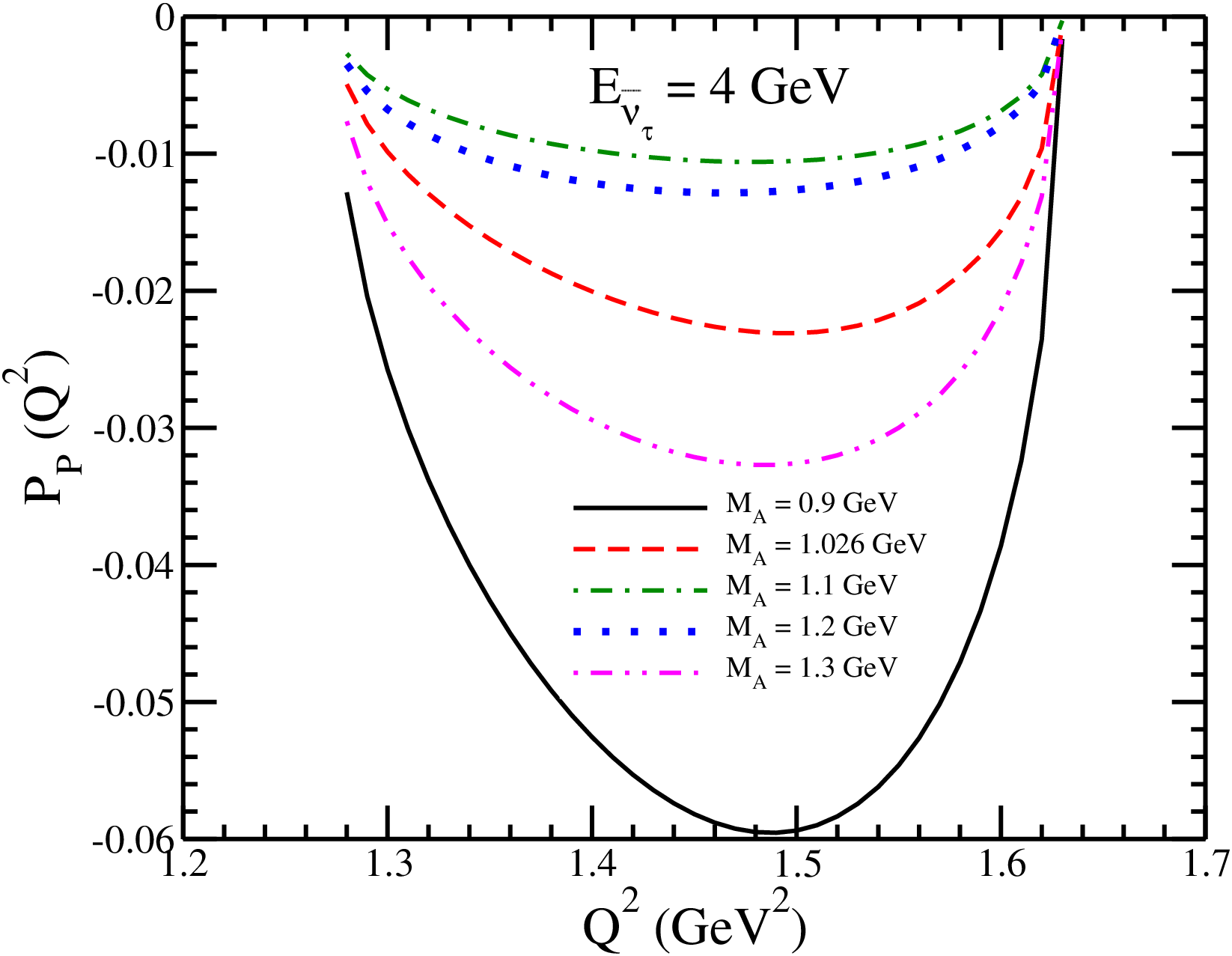}
\includegraphics[height=7cm,width=5.2cm]{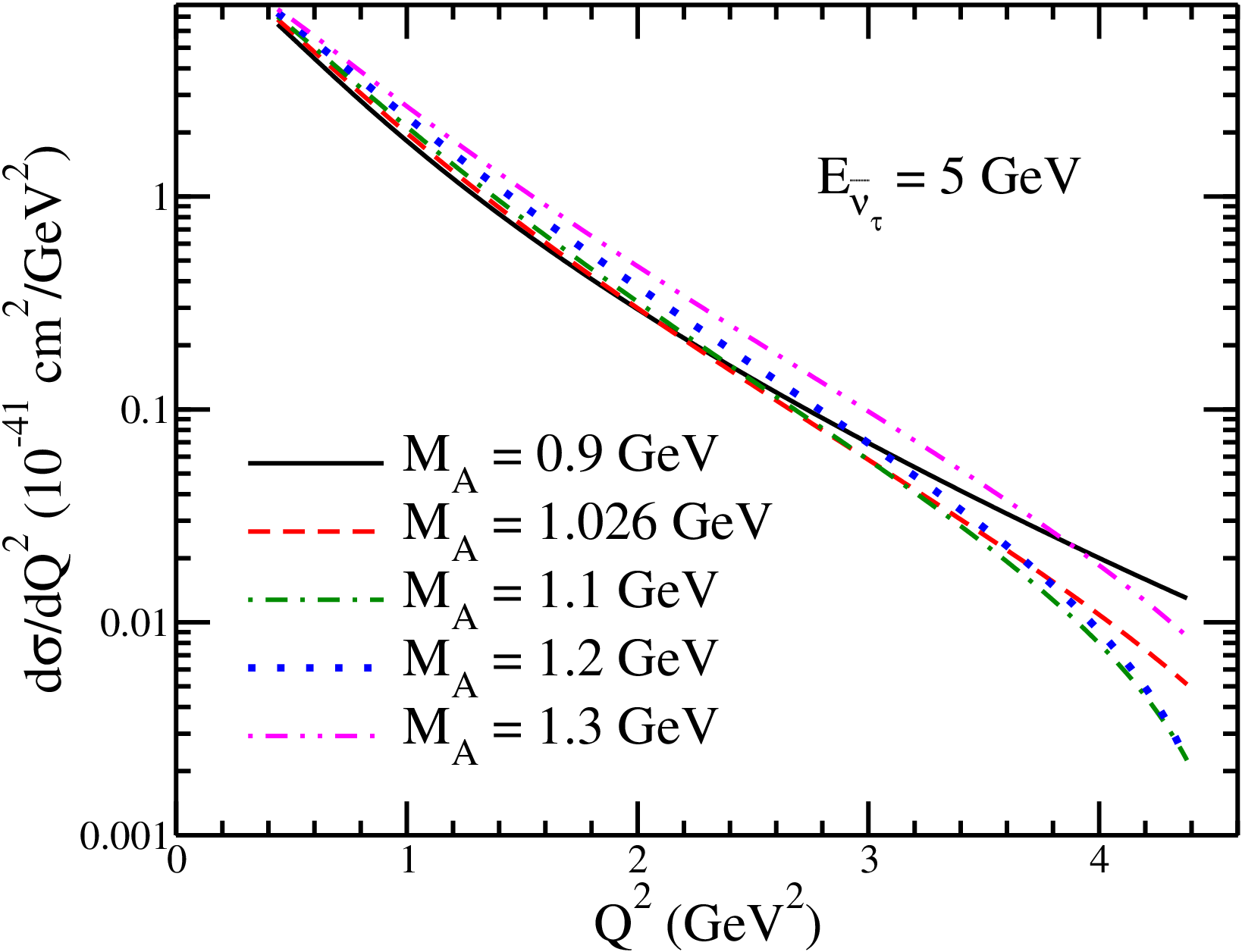}
\includegraphics[height=7cm,width=5.2cm]{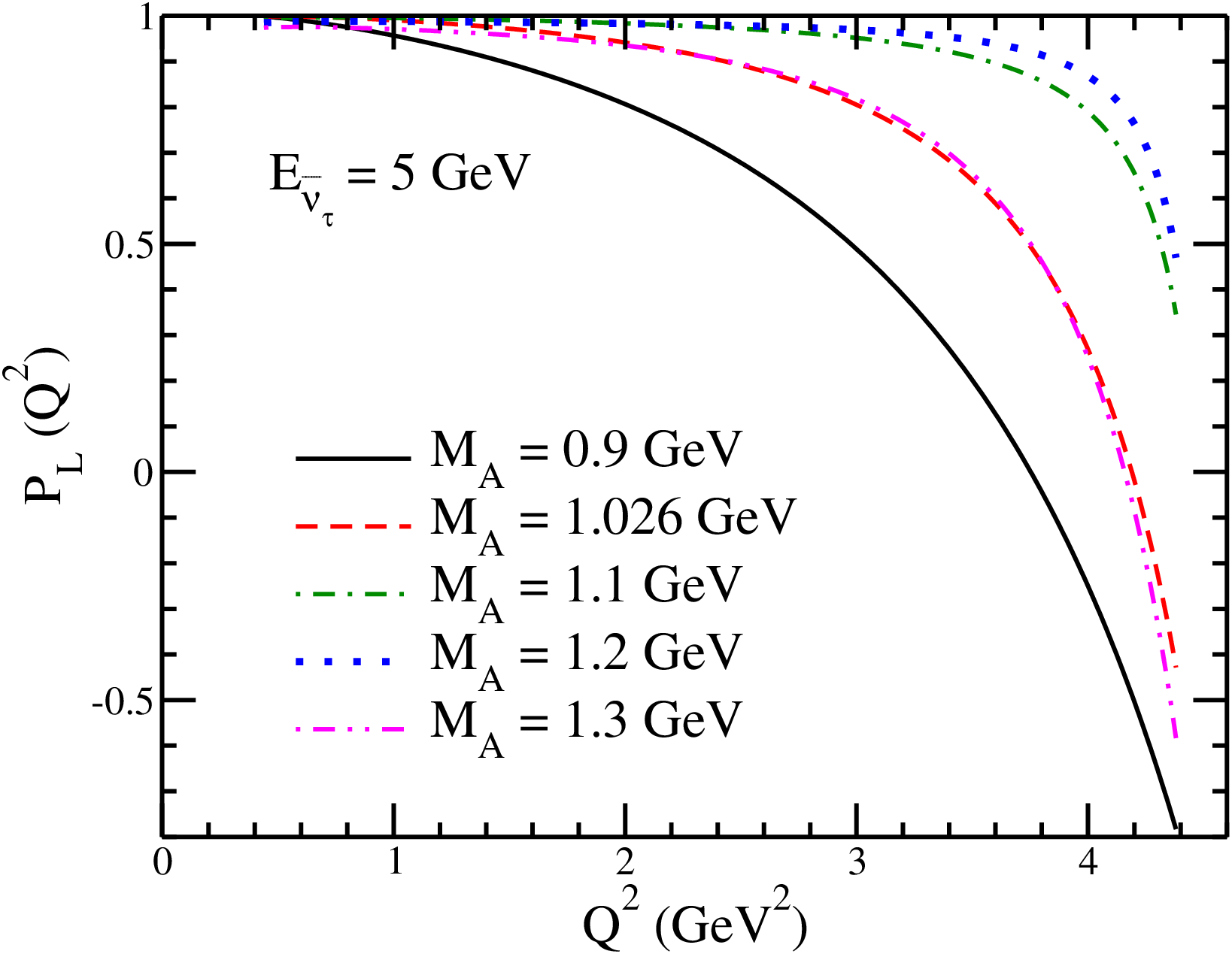}
\includegraphics[height=7cm,width=5.2cm]{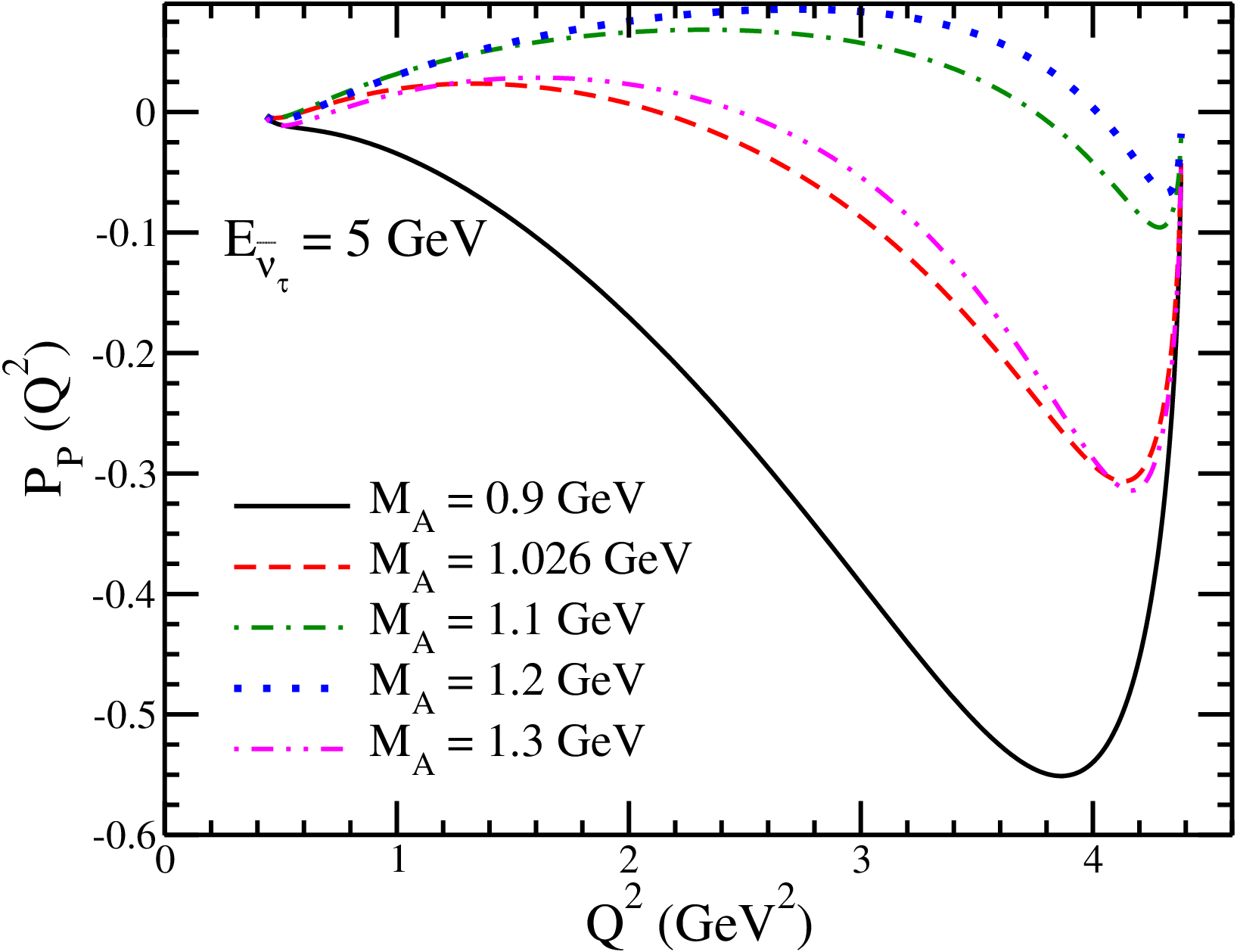}
\includegraphics[height=7cm,width=5.2cm]{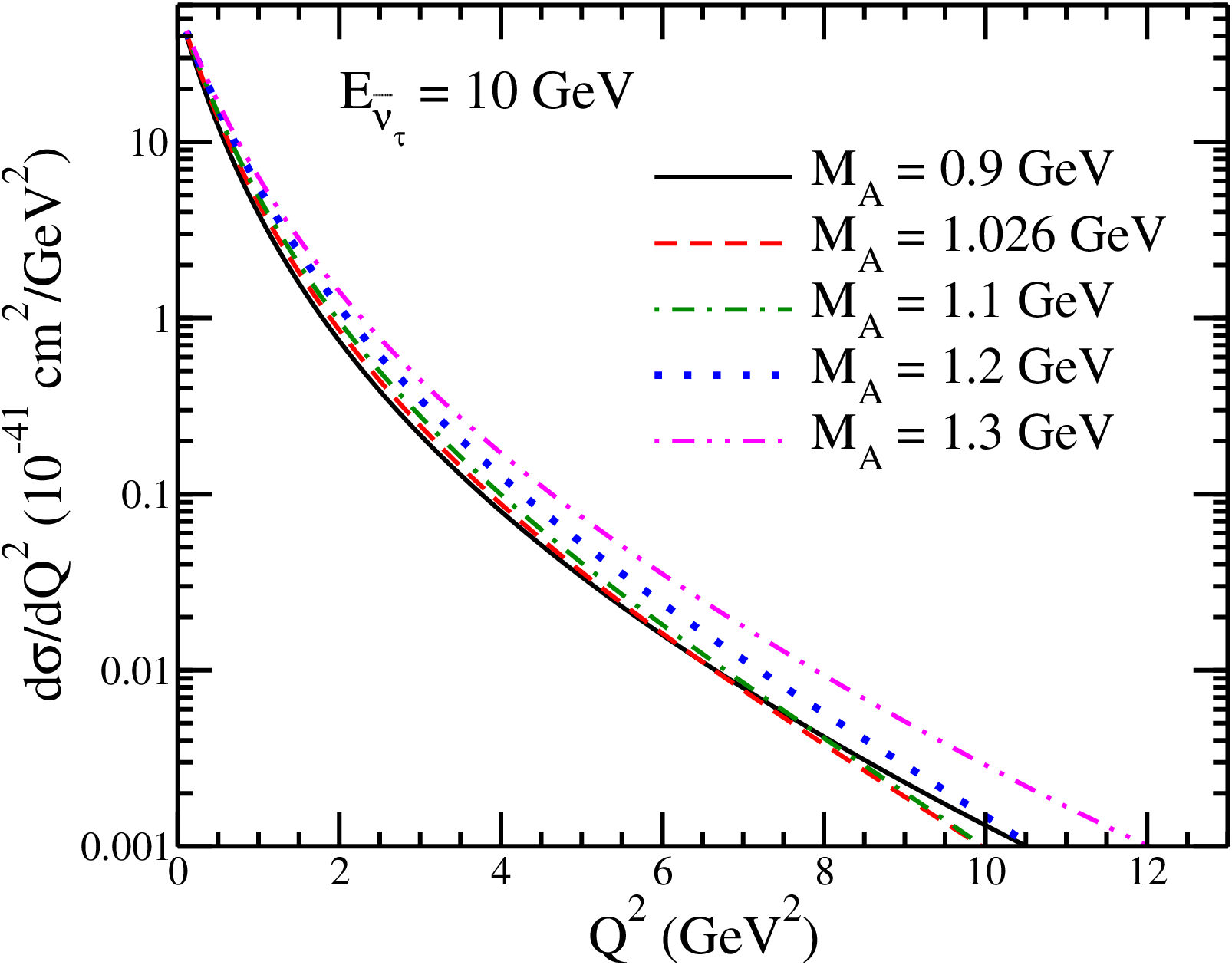} 
\includegraphics[height=7cm,width=5.2cm]{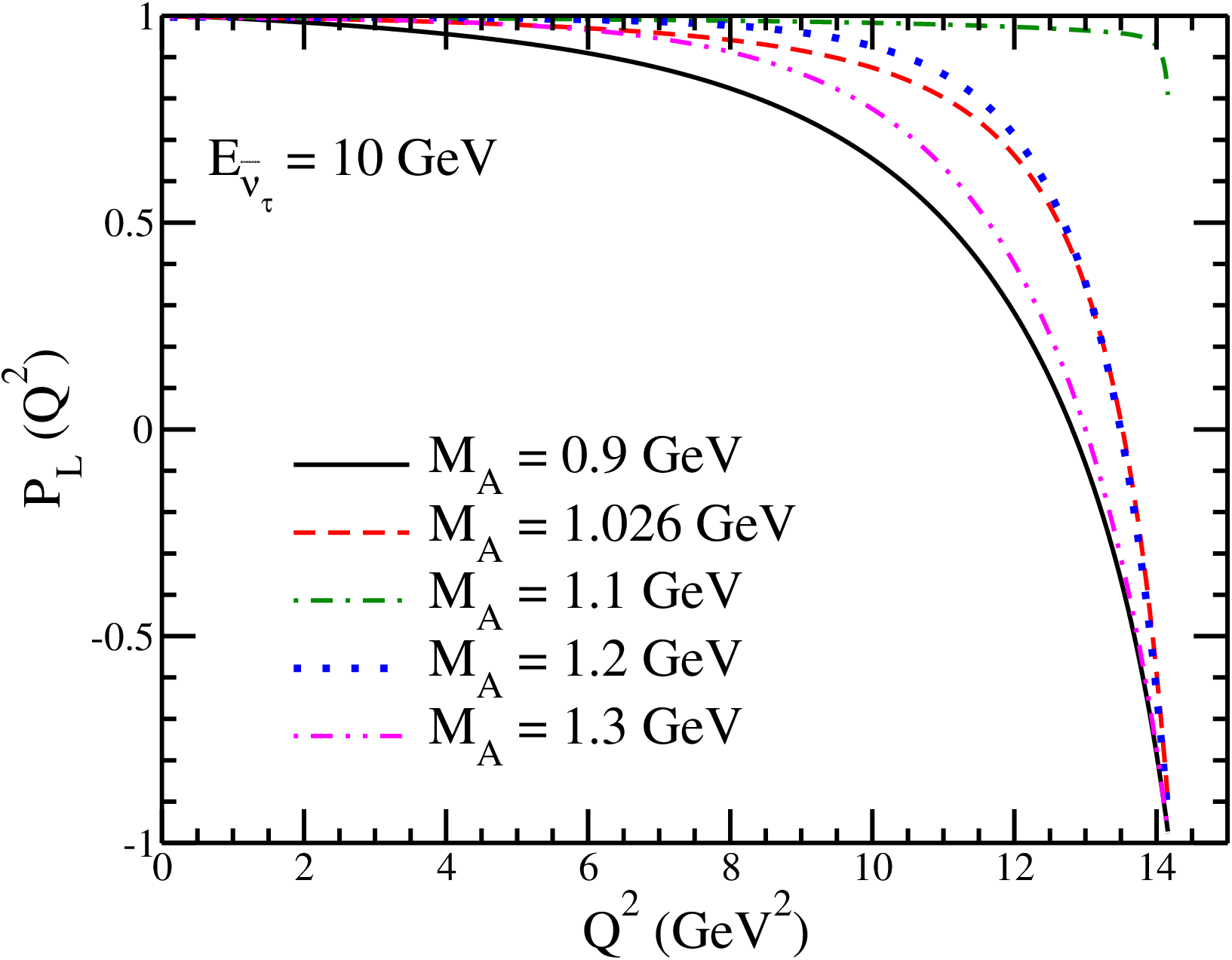}
\includegraphics[height=7cm,width=5.2cm]{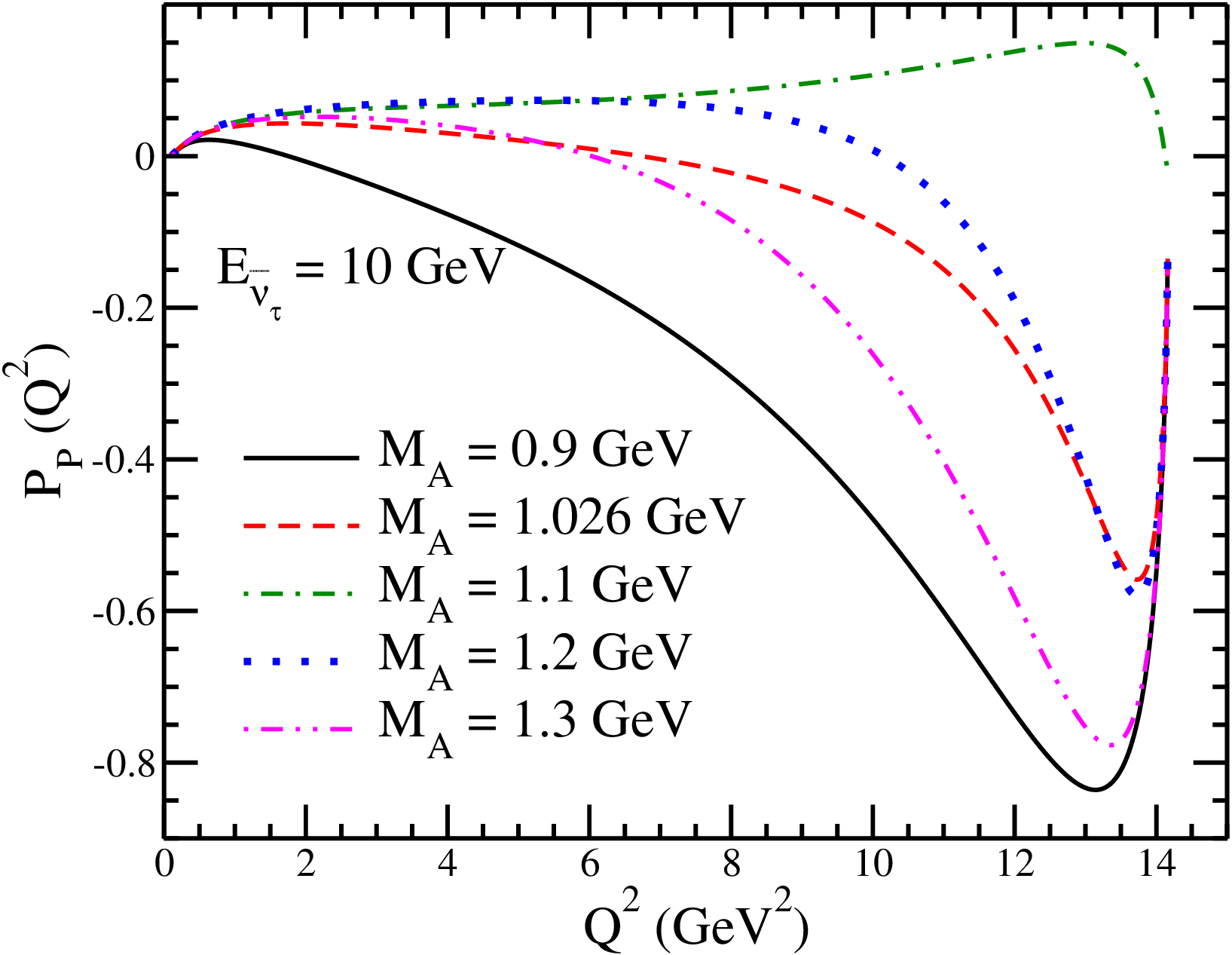}
\caption{$\frac{d\sigma}{dQ^2}$~(left panel), $P_L (Q^2)$~(middle panel) and $P_P (Q^2)$~(right panel) versus $Q^2$ for the 
process $\bar{\nu}_{\tau} + p \longrightarrow \tau^+ + \Lambda$ at $E_{\bar{\nu}_{\tau}}$ = 4 GeV~(upper panel), 5 GeV~(middle 
panel) and 10 GeV~(lower panel). The calculations have been performed using the SU(3) symmetry with the electric and magnetic 
Sachs form factors parameterized by Bradford {\it et al.}~\cite{Bradford:2006yz} and for the axial form factor, the different 
values of $M_{A}$ have been used {\it viz.} $M_{A} =$ 0.9 GeV~(solid line), 1.026 GeV~(dashed line), 1.1 GeV~(dashed-dotted 
line), 1.2 GeV~(dotted line) and 1.3 GeV~(double-dotted-dashed line).}\label{dsigma_pol_MA_lambda}
\end{figure}
\begin{figure}
\centering
\includegraphics[height=7cm,width=5.2cm]{dsig_dq2_Fp_variation_enu_4GeV_lambda_polarized.eps}
\includegraphics[height=7cm,width=5.2cm]{Pl_q2_Fp_variation_enu_4GeV_lambda.eps}
\includegraphics[height=7cm,width=5.2cm]{Pp_q2_Fp_variation_enu_4GeV_lambda.eps}
\includegraphics[height=7cm,width=5.2cm]{dsig_dq2_Fp_variation_enu_5GeV_lambda_polarized.eps}
\includegraphics[height=7cm,width=5.2cm]{Pl_q2_Fp_variation_enu_5GeV_lambda.eps}
\includegraphics[height=7cm,width=5.2cm]{Pp_q2_Fp_variation_enu_5GeV_lambda.eps}
\includegraphics[height=7cm,width=5.2cm]{dsig_dq2_Fp_variation_enu_10GeV_lambda_polarized.eps} 
\includegraphics[height=7cm,width=5.2cm]{Pl_q2_Fp_variation_enu_10GeV_lambda.eps}
\includegraphics[height=7cm,width=5.2cm]{Pp_q2_Fp_variation_enu_10GeV_lambda.eps}
\caption{$\frac{d\sigma}{dQ^2}$~(left panel), $P_L (Q^2)$~(middle panel) and $P_P (Q^2)$~(right panel) versus $Q^2$ for the 
process $\bar{\nu}_{\tau} + p \longrightarrow \tau^+ + \Lambda$ at $E_{\bar{\nu}_{\tau}}$ = 4 GeV~(upper panel), 5 GeV~(middle 
panel) and 10 GeV~(lower panel). The calculations have been performed using the SU(3) symmetry with the electric and magnetic 
Sachs form factors parameterized by Bradford {\it et al.}~\cite{Bradford:2006yz}, $M_{A}=1.026$~GeV, and for the different 
parameterizations of the pseudoscalar scalar form factor, {\it viz.} the parameterizations given by 
Nambu~\cite{Nambu:1960xd}~(solid line) and by Marshak {\it et al.}~\cite{Marshak}~(dashed line).}\label{dsigma_pol_Fp_lambda}
\end{figure}

\begin{figure}
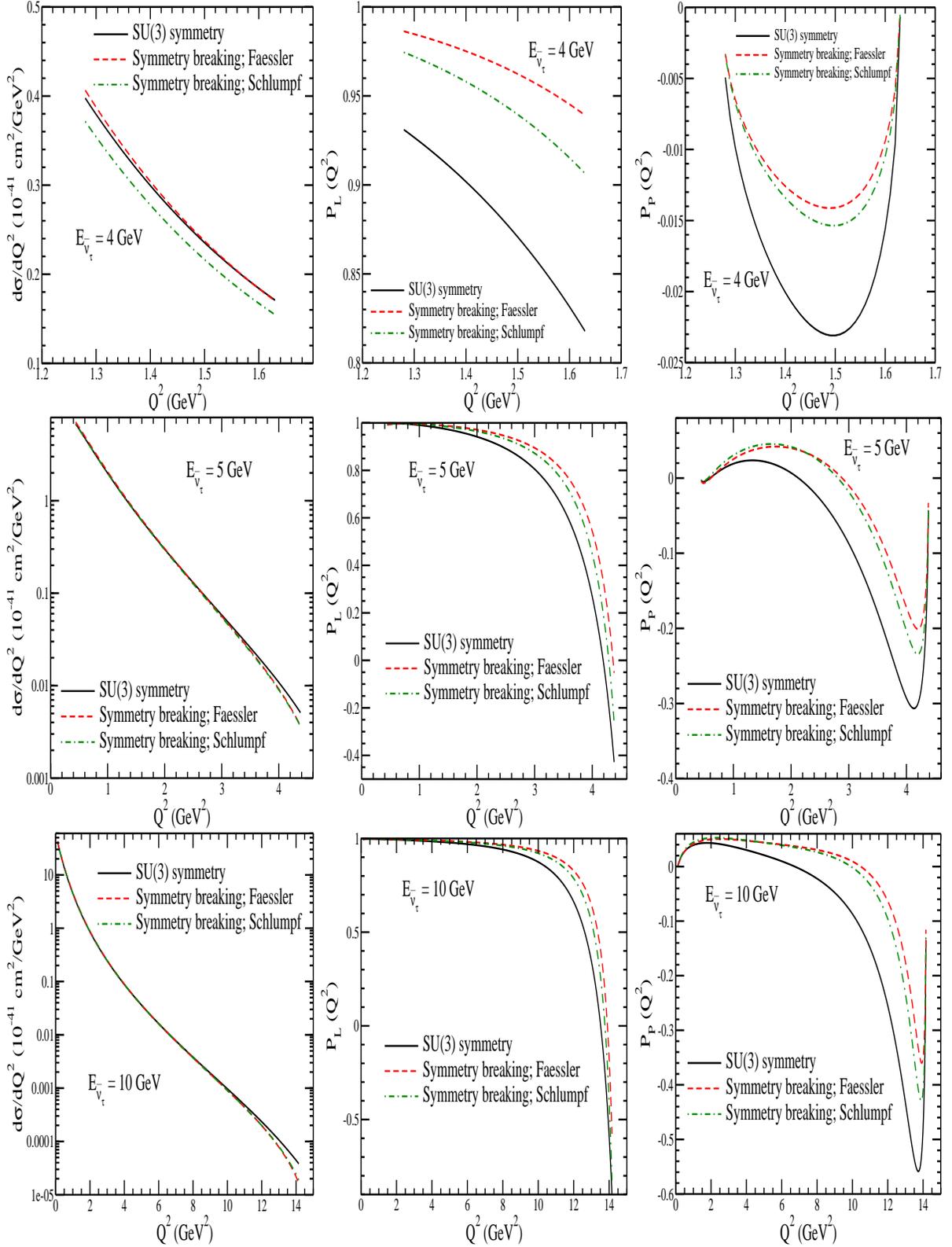

\centering
\includegraphics[height=7cm,width=5.2cm]{dsig_dq2_FF_variation_enu_4GeV_lambda_polarized_SU3.eps}
\includegraphics[height=7cm,width=5.2cm]{Pl_q2_FF_variation_enu_4GeV_lambda_SU3.eps}
\includegraphics[height=7cm,width=5.2cm]{Pp_q2_FF_variation_enu_4GeV_lambda_SU3.eps}
\includegraphics[height=7cm,width=5.2cm]{dsig_dq2_FF_variation_enu_5GeV_lambda_polarized_SU3.eps}
\includegraphics[height=7cm,width=5.2cm]{Pl_q2_FF_variation_enu_5GeV_lambda_SU3.eps}
\includegraphics[height=7cm,width=5.2cm]{Pp_q2_FF_variation_enu_5GeV_lambda_SU3.eps}
\includegraphics[height=7cm,width=5.2cm]{dsig_dq2_FF_variation_enu_10GeV_lambda_polarized_SU3.eps} 
\includegraphics[height=7cm,width=5.2cm]{Pl_q2_FF_variation_enu_10GeV_lambda_SU3.eps}
\includegraphics[height=7cm,width=5.2cm]{Pp_q2_FF_variation_enu_10GeV_lambda_SU3.eps}
\caption{$\frac{d\sigma}{dQ^2}$~(left panel), $P_L (Q^2)$~(middle panel) and $P_P (Q^2)$~(right panel) versus $Q^2$ for the 
process $\bar{\nu}_{\tau} + p \longrightarrow \tau^+ + \Lambda$ at $E_{\bar{\nu}_{\tau}}$ = 4 GeV~(upper panel), 5 GeV~(middle 
panel) and 10 GeV~(lower panel). The calculations have been performed with $M_{A} = 1.026$~GeV and 
BBBA05~\cite{Bradford:2006yz} for the vector form factors with SU(3) symmetry~(solid line), SU(3) symmetry breaking effects 
parameterized by Faessler {\it et al.}~\cite{Faessler:2008ix}~(dashed line) and the symmetry breaking effects parameterized by 
Schlumpf~\cite{Schlumpf}~(dashed-dotted line).}\label{dsigma_pol_VFF_lambda_SU3}
\end{figure}

\begin{figure}
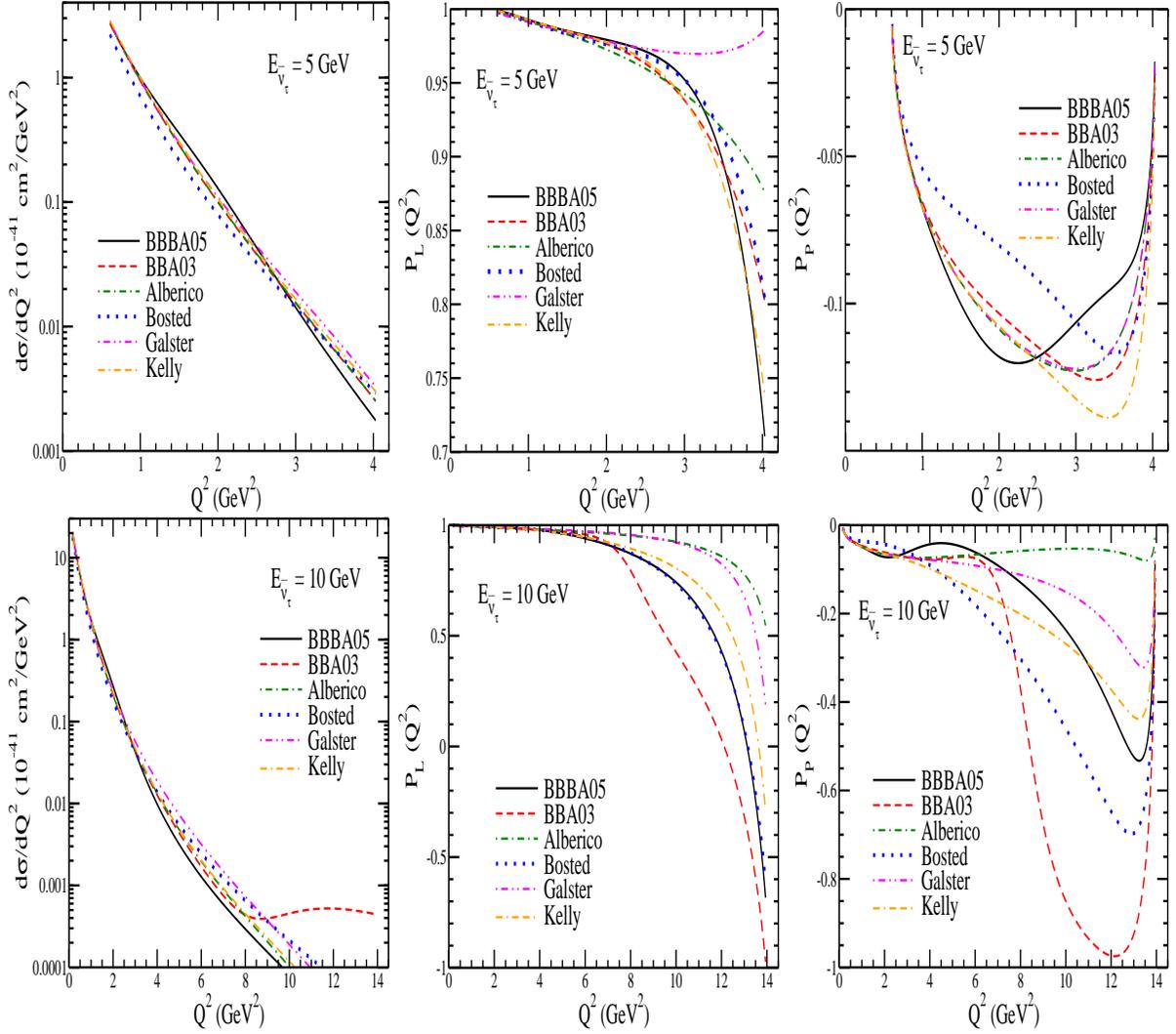

\centering
\includegraphics[height=7cm,width=5.2cm]{dsig_dq2_FF_variation_enu_5GeV_sigmaM.eps}
\includegraphics[height=7cm,width=5.2cm]{Pl_q2_FF_variation_enu_5GeV_sigmaM.eps}
\includegraphics[height=7cm,width=5.2cm]{Pp_q2_FF_variation_enu_5GeV_sigmaM.eps}
\includegraphics[height=7cm,width=5.2cm]{dsig_dq2_FF_variation_enu_10GeV_sigmaM.eps}
\includegraphics[height=7cm,width=5.2cm]{Pl_q2_FF_variation_enu_10GeV_sigmaM.eps}
\includegraphics[height=7cm,width=5.2cm]{Pp_q2_FF_variation_enu_10GeV_sigmaM.eps}
\caption{$\frac{d\sigma}{dQ^2}$~(left panel), $P_L (Q^2)$~(middle panel) and $P_P (Q^2)$~(right panel) versus $Q^2$ for the 
process $\bar{\nu}_{\tau} + n \longrightarrow \tau^+ + \Sigma^{-}$ at $E_{\bar{\nu}_{\tau}}$ = 5 GeV~(upper panel) and 10 
GeV~(lower panel). The calculations have been performed using the SU(3) symmetry with the axial dipole mass $M_{A} = 
1.026$~GeV and for the different parameterizations of the nucleon vector form factors {\it viz.}, 
BBBA05~\cite{Bradford:2006yz}~(solid line), BBA03~\cite{Budd:2004bp}~(dashed line), Alberico {\it et 
al.}~\cite{Alberico:2008sz}~(dashed-dotted line), Bosted~\cite{Bosted:1994tm}~(dotted line), Galster {\it et 
al.}~\cite{Galster:1971kv}~(double-dotted-dashed line) and Kelly~\cite{Kelly:2004hm}~(double-dashed-dotted line).}
\label{dsigma_pol_VFF_sigma}
\end{figure}

\begin{figure}
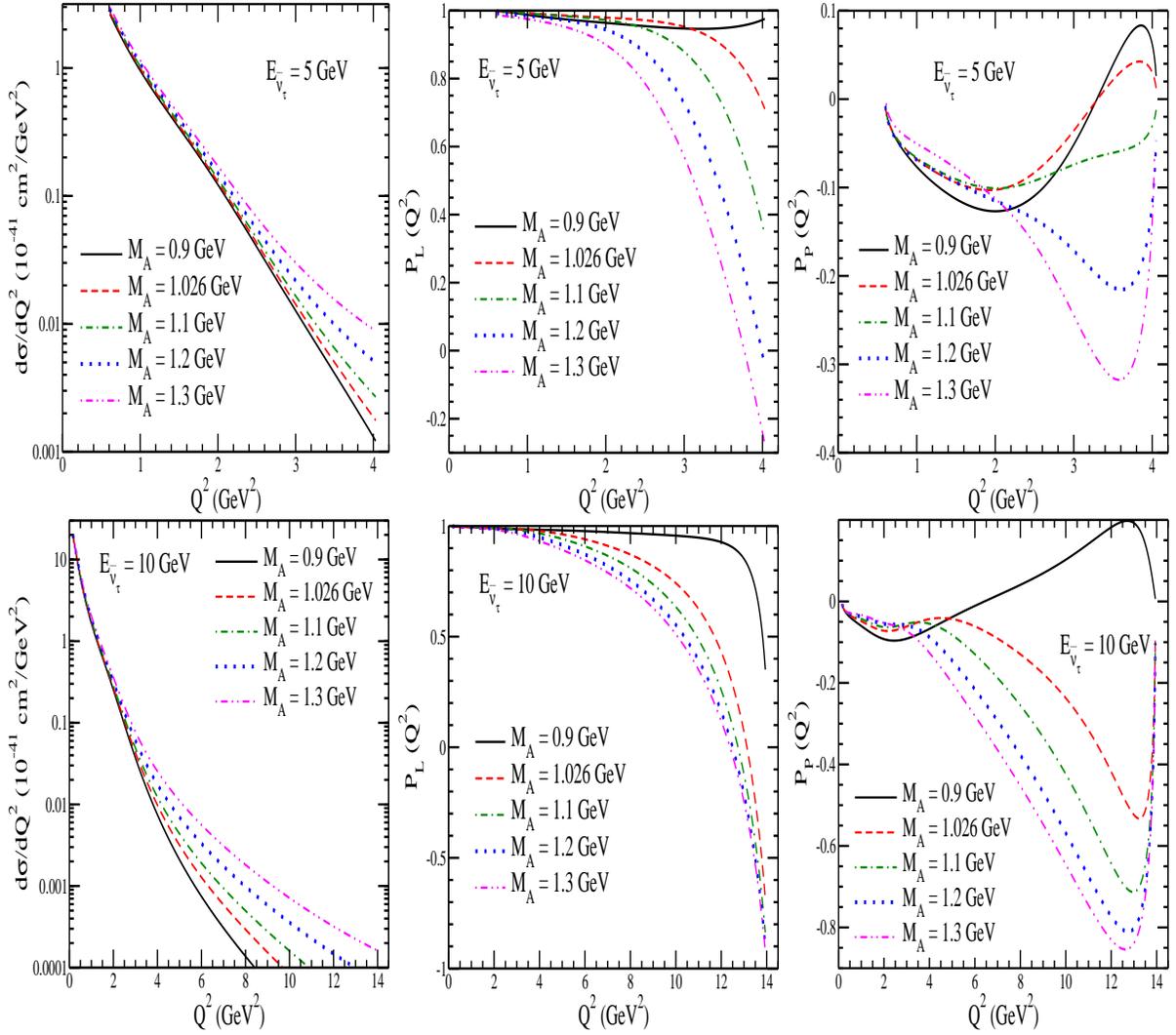

\centering
\includegraphics[height=7cm,width=5.2cm]{dsig_dq2_Ma_variation_enu_5GeV_sigmaM.eps}
\includegraphics[height=7cm,width=5.2cm]{Pl_q2_Ma_variation_enu_5GeV_sigmaM.eps}
\includegraphics[height=7cm,width=5.2cm]{Pp_q2_Ma_variation_enu_5GeV_sigmaM.eps}
\includegraphics[height=7cm,width=5.2cm]{dsig_dq2_Ma_variation_enu_10GeV_sigmaM.eps}
\includegraphics[height=7cm,width=5.2cm]{Pl_q2_Ma_variation_enu_10GeV_sigmaM.eps}
\includegraphics[height=7cm,width=5.2cm]{Pp_q2_Ma_variation_enu_10GeV_sigmaM.eps}
\caption{$\frac{d\sigma}{dQ^2}$~(left panel), $P_L (Q^2)$~(middle panel) and $P_P (Q^2)$~(right panel) versus $Q^2$ for the 
process $\bar{\nu}_{\tau} + n \longrightarrow \tau^+ + \Sigma^{-}$ at $E_{\bar{\nu}_{\tau}}$ = 5 GeV~(upper panel) and 10 
GeV~(lower panel). The calculations have been performed using the SU(3) symmetry with the electric and magnetic Sachs form 
factors parameterized by Bradford {\it et al.}~\cite{Bradford:2006yz} and for the axial form factor, the different values 
of $M_{A}$ have been used {\it viz.} $M_{A} =$ 0.9 GeV~(solid line), 1.026 GeV~(dashed line), 1.1 GeV~(dashed-dotted line), 
1.2 GeV~(dotted line) and 1.3 GeV~(double-dotted-dashed line).}\label{dsigma_pol_MA_sigma}
\end{figure}

\begin{figure}
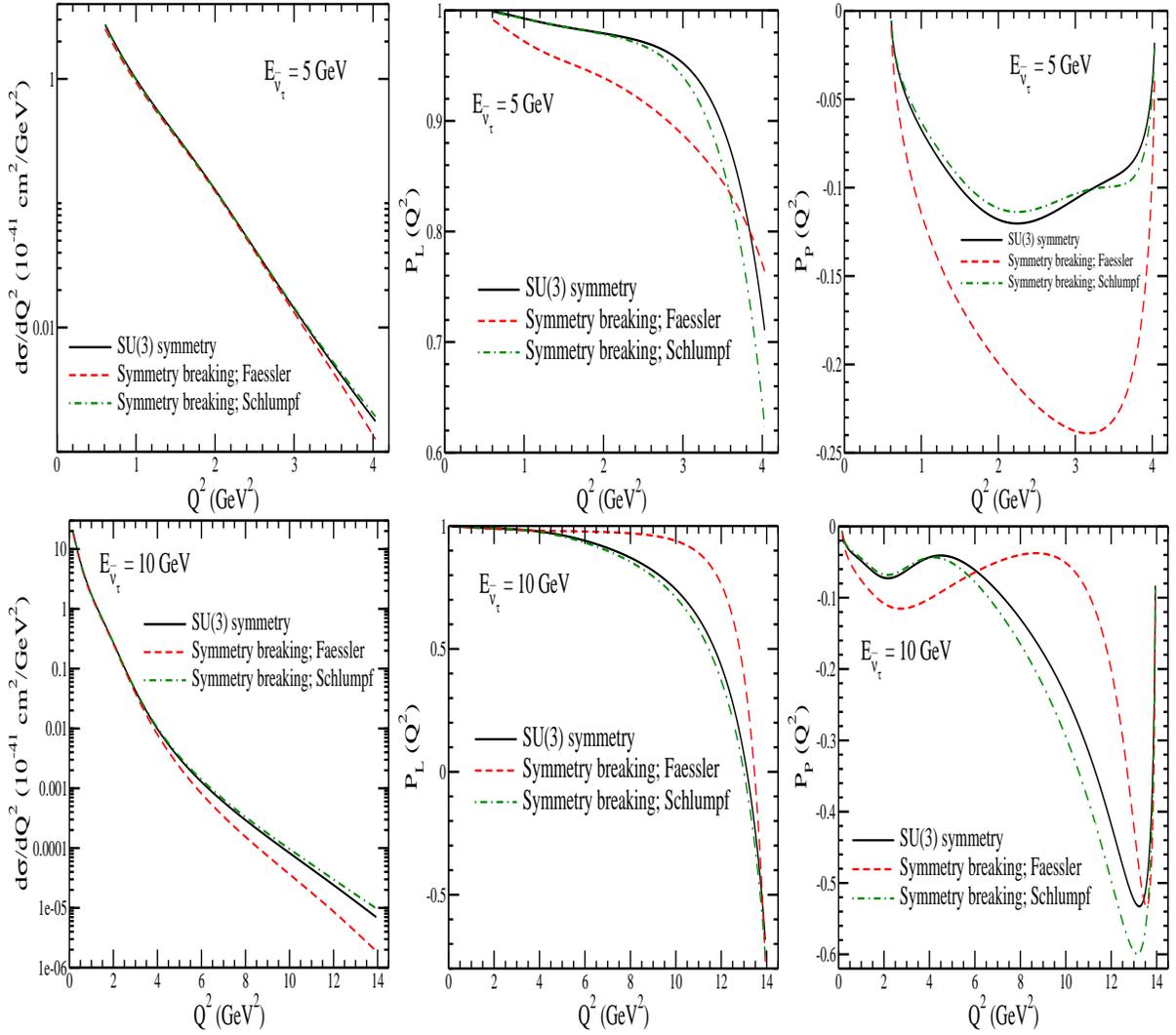

\centering
\includegraphics[height=7cm,width=5.2cm]{dsig_dq2_Ma_variation_enu_5GeV_sigmaM_SU3.eps}
\includegraphics[height=7cm,width=5.2cm]{Pl_q2_Ma_variation_enu_5GeV_sigmaM_SU3.eps}
\includegraphics[height=7cm,width=5.2cm]{Pp_q2_Ma_variation_enu_5GeV_sigmaM_SU3.eps}
\includegraphics[height=7cm,width=5.2cm]{dsig_dq2_Ma_variation_enu_10GeV_sigmaM_SU3.eps}
\includegraphics[height=7cm,width=5.2cm]{Pl_q2_Ma_variation_enu_10GeV_sigmaM_SU3.eps}
\includegraphics[height=7cm,width=5.2cm]{Pp_q2_Ma_variation_enu_10GeV_sigmaM_SU3.eps}
\caption{$\frac{d\sigma}{dQ^2}$~(left panel), $P_L (Q^2)$~(middle panel) and $P_P (Q^2)$~(right panel) versus $Q^2$ for the 
process $\bar{\nu}_{\tau} + n \longrightarrow \tau^+ + \Sigma^{-}$ at $E_{\bar{\nu}_{\tau}}$ = 5 GeV~(upper panel) and 10 
GeV~(lower panel). The calculations have been performed with $M_{A} = 1.026$~GeV assuming SU(3) symmetry~(solid line), 
as well as with SU(3) symmetry breaking effects parameterized by Faessler {\it et al.}~\cite{Faessler:2008ix}~(dashed line) 
and by Schlumpf~\cite{Schlumpf}~(dashed-dotted line).}\label{dsigma_pol_MA_sigma_SU3}
\end{figure}

We have used Eqs.~(\ref{dsig}), ~(\ref{Pll}) and (\ref{Ppl}), respectively, to numerically evaluate the differential 
scattering cross section $d \sigma/d Q^2$, the longitudinal~($P_L(Q^2)$) and the perpendicular~($P_P(Q^2)$) components of 
polarization of $\tau$ lepton. The Dirac and Pauli form factors $f_{1,2}^{N} (Q^2); ~ (N=p,n)$ are expressed in terms of the 
electric and magnetic Sachs' form factors, for which the various parameterizations~\cite{Bradford:2006yz, Budd:2004bp, 
Bosted:1994tm, Alberico:2008sz, Kelly:2004hm, Galster:1971kv} available in the literature, have been used. For $g_1 (Q^2)$ a 
dipole parameterization has been used~(Eq.~(\ref{ga})), with the world average value of the axial dipole mass $M_A = $ 1.026 
GeV. For the pseudoscalar form factor $g_3 (Q^2)$, the parameterizations given by Nambu~\cite{Nambu:1960xd} and Marshak 
{\it et al.}~\cite{Marshak} have been used. The numerical results of the differential scattering cross section and 
polarization observables obtained assuming SU(3) symmetry and the results with SU(3) symmetry breaking effects, using the 
prescriptions of: (i) Faessler {\it et al.}~\cite{Faessler:2008ix}, and (ii) Schlumpf~\cite{Schlumpf}, are presented 
separately for the $\Lambda$ and $\Sigma^{-}$ productions from the nucleons. The results for the $\Sigma^{0}$ 
production can be expressed in terms of the $\Sigma^{-}$ production in the SU(3) symmetric limit~(Eqs.~(\ref{sigma_sigma} and 
\ref{P_sigma})) and are not presented separately.

\subsubsection{$\Lambda$ production}
In Fig.~\ref{dsigma_pol_VFF_lambda}, we present the results for the $Q^2$ distribution {\it i.e.} $\frac{d\sigma}{dQ^2}$,
$P_L (Q^2)$ and $P_P (Q^2)$ vs $Q^2$ for $\bar{\nu}_{\tau} + p \longrightarrow \tau^+ + \Lambda$ process at the three 
different values of energy viz. $E_{\bar{\nu}_{\tau}} =$ 4 GeV, 5 GeV and 10 GeV, assuming SU(3) symmetry with $M_{A} = 
1.026$~GeV and using the different parameterizations of the nucleon vector form factors {\it viz.}, 
BBBA05~\cite{Bradford:2006yz}, BBA03~\cite{Budd:2004bp}, Alberico~\cite{Alberico:2008sz}, Bosted~\cite{Bosted:1994tm}, 
Galster~\cite{Galster:1971kv} and Kelly~\cite{Kelly:2004hm}. We see that at low ${\bar\nu}_\tau$ energies, for example at 
$E_{\bar{\nu}_{\tau}} =4$~GeV, there is considerable dependence of the different parameterizations of the vector form factors 
on $\frac{d\sigma}{dQ^2}$, $P_L (Q^2)$ and $P_P (Q^2)$ distributions. However, with the increase in ${\bar\nu}_\tau$ energy, 
this difference decreases and becomes almost negligible at higher energies, like at $E_{\bar{\nu}_{\tau}} =10$~GeV, 
especially for $\frac{d\sigma}{dQ^2}$ and $P_L (Q^2)$ distributions. 

To study the effect of the variation in $M_{A}$~(in the range 0.9--1.3~GeV) on the differential cross section and polarization 
observables, we present, in Fig.~\ref{dsigma_pol_MA_lambda}, the results for $\frac{d\sigma}{dQ^2}$, $P_L (Q^2)$ and 
$P_P (Q^2)$ distributions at $E_{\bar{\nu}_{\tau}}$ = 4 GeV, 5 GeV and 10 GeV. We find that at low ${\bar\nu}_\tau$ 
energies, there is a significant dependence of these distributions on the choice of $M_A$. With the increase in 
${\bar\nu}_\tau$ energy, this dependence on the variation in $M_A$ decreases, especially for $\frac{d\sigma}{dQ^2}$ and to 
some extent for $P_L (Q^2)$ but not for $P_P (Q^2)$ distribution. Moreover, it is important to point out that in the case of 
$\bar{\nu}_{\tau} + p \longrightarrow \tau^{+} + \Lambda$ reaction, with the increase in $M_{A}$, $\frac{d\sigma}{dQ^2}$ 
decreases~(0.9 GeV to 1.1 GeV), but with the further increase in $M_{A}$~(1.1 GeV to 1.3 GeV), $\frac{d\sigma}{dQ^2}$ 
increases, which is not generally the case in ${\nu}_{l} + n \longrightarrow l^{-} + p; (l=e,\mu,\tau)$ scattering. Moreover, 
in the case of $\bar{\nu}_{l} + p \longrightarrow l^{+} + n$, we have observed that with the increase in $M_{A}$, 
$\frac{d\sigma}{dQ^2}$ decreases~(from 0.9 GeV to 1.1 GeV) and with further increase in $M_{A}=1.2$~GeV, $\frac{d\sigma}
{dQ^2}$ increases~\cite{Fatima:2020pvv}. In the present work, for $\Lambda$ production we observe similar trend as in the 
case of $\bar{\nu}_{\tau}$ induced CCQE reaction~\cite{Fatima:2020pvv}. In the $\bar{\nu}_{\tau}$ induced reactions because 
of the production of massive $\tau$ lepton in the final state, the pseudoscalar form factor becomes significant. The variation 
in $\frac{d\sigma}{dQ^2}$ observed in Fig.~\ref{dsigma_pol_MA_lambda} arises due to interference of the pseudoscalar form 
factor with axial vector and vector form factors.

\begin{figure}
\centering
\includegraphics[height=7cm,width=5.2cm]{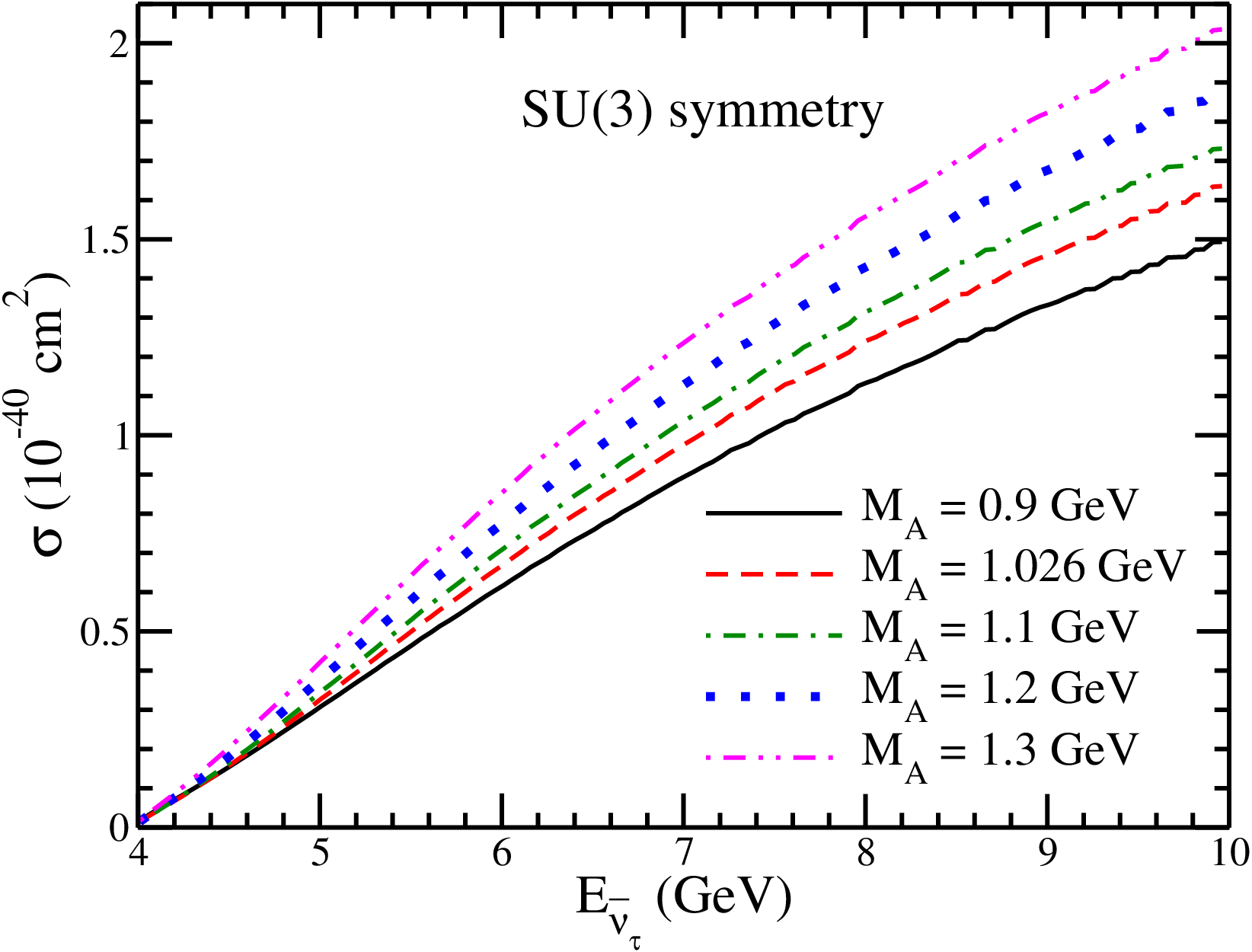}
\includegraphics[height=7cm,width=5.2cm]{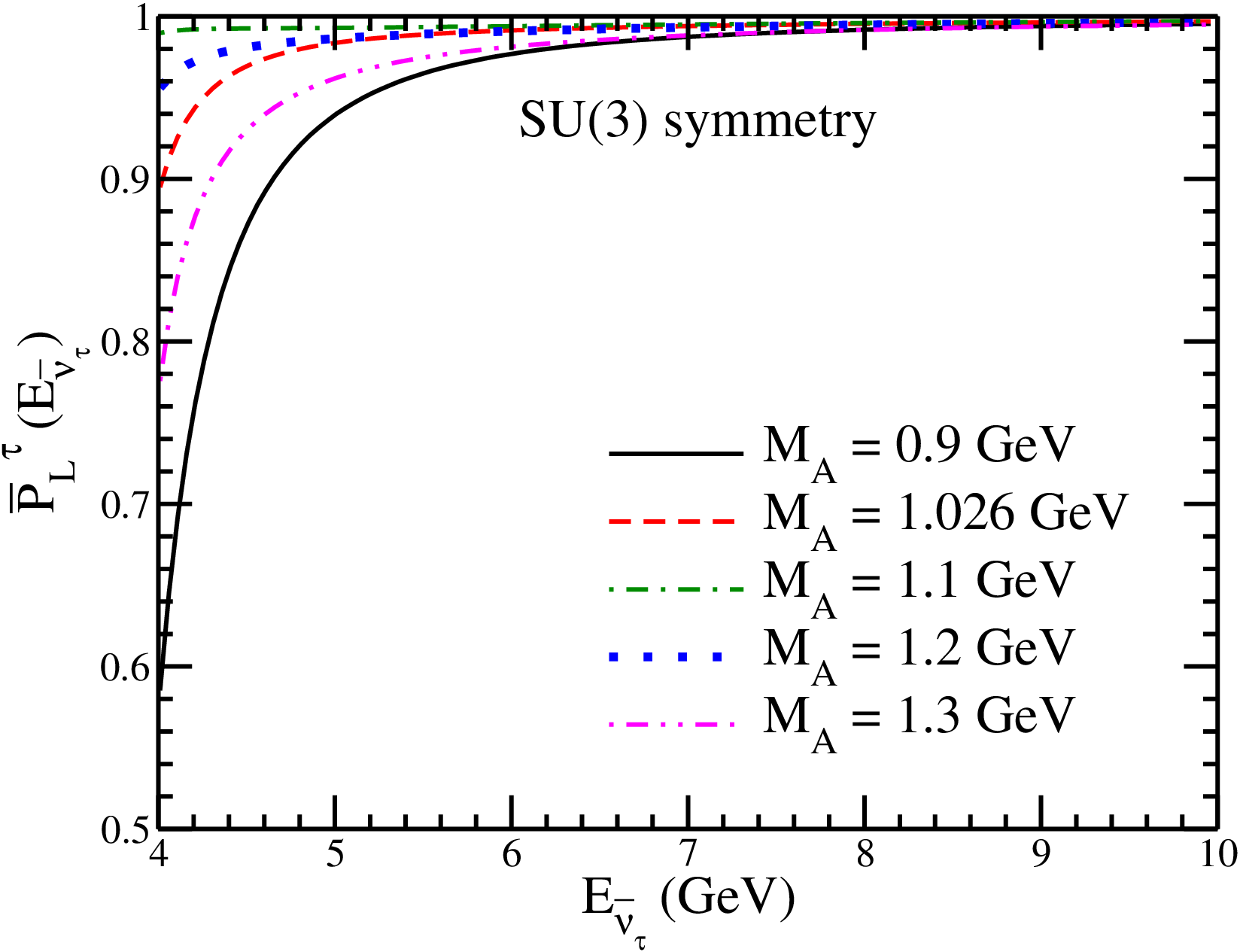}
\includegraphics[height=7cm,width=5.2cm]{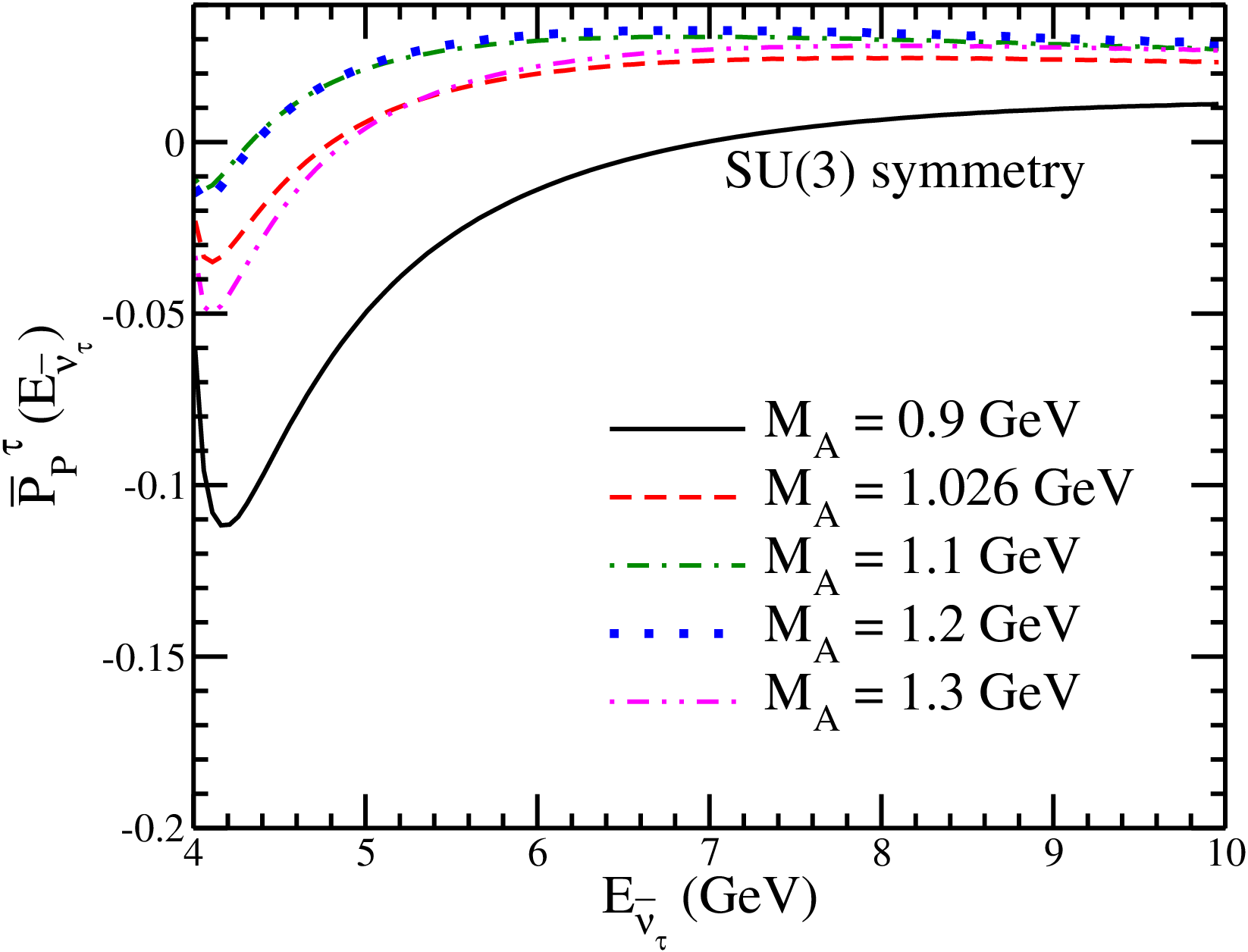}
\caption{$\sigma$~(left panel), $\overline{P}_{L} (E_{\bar{\nu}_{\tau}})$~(middle panel), and $\overline{P}_{P} 
(E_{\bar{\nu}_{\tau}})$~(right panel) vs $E_{\bar{\nu}_{\tau}}$ for $\bar{\nu}_{\tau} + p \rightarrow \tau^{+} + \Lambda$ 
process. The calculations have been performed using the SU(3) symmetry with electric and magnetic Sachs form factors 
parameterized by Bradford {\it et al.}~\cite{Bradford:2006yz} and for the axial form factor, the different values of $M_{A}$ 
have been used {\it viz.} $M_{A} =$ 0.9 GeV~(solid line), 1.026 GeV~(dashed line), 1.1 GeV~(dashed-dotted line), 1.2 
GeV~(dotted line) and 1.3 GeV~(double-dotted-dashed line).}\label{sigma_MA_lambda}
\end{figure}

To see the dependence of $\frac{d\sigma}{dQ^2}$, $P_L (Q^2)$ and $P_P (Q^2)$ on the pseudoscalar form factor $g_3^{NY} 
(Q^2)$, we have used the two parameterizations of $g_{3}^{NY} (Q^2)$ given in Eqs.~(\ref{g3}) by Nambu~\cite{Nambu:1960xd}, 
and (\ref{g3_Marshak}) by Marshak {\it et al.}~\cite{Marshak}, and show the numerical results in 
Fig.~\ref{dsigma_pol_Fp_lambda}. It may be observed that at low ${\bar\nu}_\tau$ energies, there is a large dependence on the 
choice of $g_3^{NY}(Q^2)$. While with the increase in $E_{\bar{\nu}_{\tau}}$, this dependence on the choice of $g_3^{NY} 
(Q^2)$ becomes almost negligible for the $\frac{d\sigma}{dQ^2}$ distribution, whereas for $P_L (Q^2)$ and $P_P (Q^2)$ 
distributions, they are found to be quite significant, even for the higher values of $E_{\bar{\nu}_{\tau}}$, say 
$E_{\bar{\nu}_{\tau}}=10$~GeV.

In Fig.~\ref{dsigma_pol_VFF_lambda_SU3}, we present the results for $\frac{d\sigma}{dQ^2}$, $P_L (Q^2)$ and $P_P (Q^2)$ vs 
$Q^2$ at $E_{\bar{\nu}_{\tau}}$ = 4 GeV, 5 GeV and 10 GeV, when the SU(3) symmetry breaking effects are taken into account 
following the prescription of Faessler {\it et al.}~\cite{Faessler:2008ix} and Schlumpf~\cite{Schlumpf}. We observe that at 
low ${\bar\nu}_\tau$ energies, for example at $E_{\bar{\nu}_{\tau}} =4$~GeV, there is some effect of SU(3) symmetry breaking 
on the $\frac{d\sigma}{dQ^2}$ distribution if the parameterization by Schlumpf~\cite{Schlumpf} is used, while there is almost 
no effect if one uses the parameterization of Faessler {\it et al.}~\cite{Faessler:2008ix}. However, in the case of 
$P_L (Q^2)$ and $P_P (Q^2)$, there is some effect of SU(3) breaking using either the prescription of Faessler {\it et 
al.}~\cite{Faessler:2008ix} or Schlumpf~\cite{Schlumpf}. Moreover, with the increase in ${\bar\nu}_\tau$ energy, the 
difference in $\frac{d\sigma}{dQ^2}$ due to the use of the two parameterizations of SU(3) symmetry breaking becomes almost 
negligible at higher energies. In the case of polarization observables, the two parameterizations of SU(3) symmetry breaking 
give different results for all the values of $Q^2$. 

\subsubsection{$\Sigma^{-}$ production}
In Fig.~\ref{dsigma_pol_VFF_sigma}, we present the results for $\frac{d\sigma}{dQ^2}$, $P_L (Q^2)$ and $P_P (Q^2)$ vs $Q^2$ 
for the process $\bar{\nu}_{\tau} + n \longrightarrow \tau^+ + \Sigma^-$ at the two different value of energies viz. 
$E_{\bar{\nu}_{\tau}}$ = 5 GeV and 10 GeV, using the different parameterizations of the nucleon vector form factors {\it 
viz.}, BBBA05~\cite{Bradford:2006yz}, BBA03~\cite{Budd:2004bp}, Alberico~\cite{Alberico:2008sz}, Bosted~\cite{Bosted:1994tm}, 
Galster~\cite{Galster:1971kv} and Kelly~\cite{Kelly:2004hm}. We observe that at low ${\bar\nu}_\tau$ energies, for example 
at $E_{\bar{\nu}_{\tau}}=5$~GeV, there is considerable dependence of the different parameterizations of the vector form 
factors on $\frac{d\sigma}{dQ^2}$, $P_L (Q^2)$ and $P_P (Q^2)$ distributions. However, unlike the case of $\Lambda$ 
production~(Fig.~\ref{dsigma_pol_MA_lambda}), with the increase in ${\bar\nu}_\tau$ energy, this difference further increases 
and becomes quite significant at higher energies, especially for the distributions of the polarization observables 
$P_L (Q^2)$ and $P_P (Q^2)$. 

The effect of $M_{A}$ variation on the differential cross section and polarization observables are presented in 
Fig.~\ref{dsigma_pol_MA_sigma}, at $E_{\bar{\nu}_{\tau}}$ = 5 GeV and 10 GeV by varying $M_{A}$ in the range 0.9--1.3~GeV. 
We find that at low ${\bar\nu}_\tau$ energies, there is large dependence of these distributions on the choice of $M_A$, which 
increases with increase in antineutrino energy. 

In Fig.~\ref{dsigma_pol_MA_sigma_SU3}, we show the effect of SU(3) symmetry breaking by taking into account the 
parameterizations given by Faessler {\it et al.}~\cite{Faessler:2008ix} and Schlumpf~\cite{Schlumpf} on $\frac{d\sigma}
{dQ^2}$, $P_L (Q^2)$ and $P_P (Q^2)$ vs $Q^2$ at $E_{\bar{\nu}_{\tau}}$ = 5 GeV and 10 GeV. From the figure, it may be 
observed that in the case of $\frac{d\sigma}{dQ^2}$, at low energies, say $E_{\bar{\nu}_{\tau}} = 5$~GeV, the effect of 
SU(3) symmetry breaking is quite small. While at higher energies, say $E_{\bar{\nu}_{\tau}}=10$~GeV, the effect of SU(3) 
breaking is seen in the model of Faessler {\it et al.}~\cite{Faessler:2008ix} but not in the model of 
Schlumpf~\cite{Schlumpf}. Whereas the polarization observables~(both $P_{L} (Q^2)$ and $P_{P} (Q^2)$) are quite sensitive to 
SU(3) symmetry breaking effect, and there is considerable change in the results obtained using the two different choices of 
parameterizing the SU(3) symmetry breaking effect.

\begin{figure}
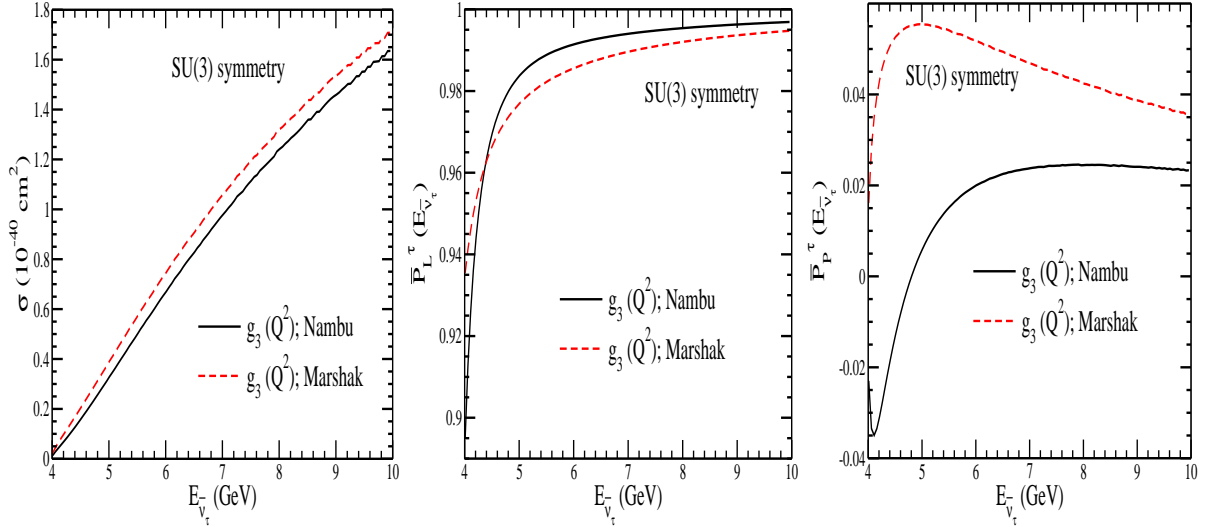

\centering
\includegraphics[height=7cm,width=5.2cm]{sigma_FP_variation_lambda.eps}
\includegraphics[height=7cm,width=5.2cm]{Pl_FP_variation_lambda.eps}
\includegraphics[height=7cm,width=5.2cm]{Pp_FP_variation_lambda.eps}
\caption{$\sigma$~(left panel), $\overline{P}_{L} (E_{\bar{\nu}_{\tau}})$~(middle panel), and $\overline{P}_{P} 
(E_{\bar{\nu}_{\tau}})$~(right panel) vs $E_{\bar{\nu}_{\tau}}$ for $\bar{\nu}_{\tau} + p \rightarrow \tau^{+} + \Lambda$ 
process. The calculations have been performed using the SU(3) symmetry with electric and magnetic Sachs form factors 
parameterized by Bradford {\it et al.}~\cite{Bradford:2006yz} and for the axial form factor, $M_{A} = 1.026$~GeV is used, 
with the different parameterizations of the pseudoscalar form factor {\it viz.}, using the parameterizations given by 
Nambu~\cite{Nambu:1960xd}~(solid line) and Marshak {\it et al.}~\cite{Marshak}~(dashed line).}\label{sigma_FP_lambda}
\end{figure}

\begin{figure}
\centering
\includegraphics[height=7cm,width=5.2cm]{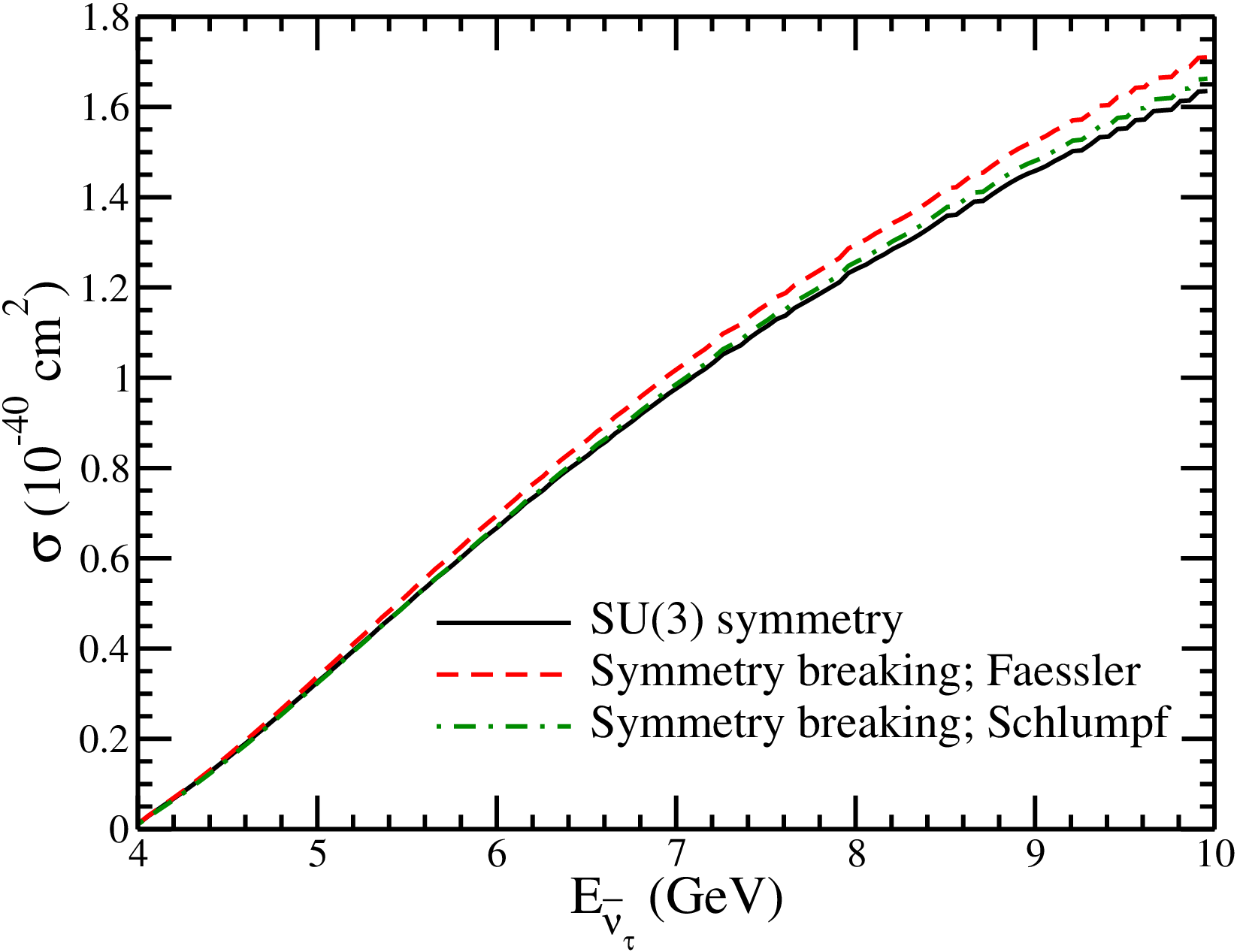}
\includegraphics[height=7cm,width=5.2cm]{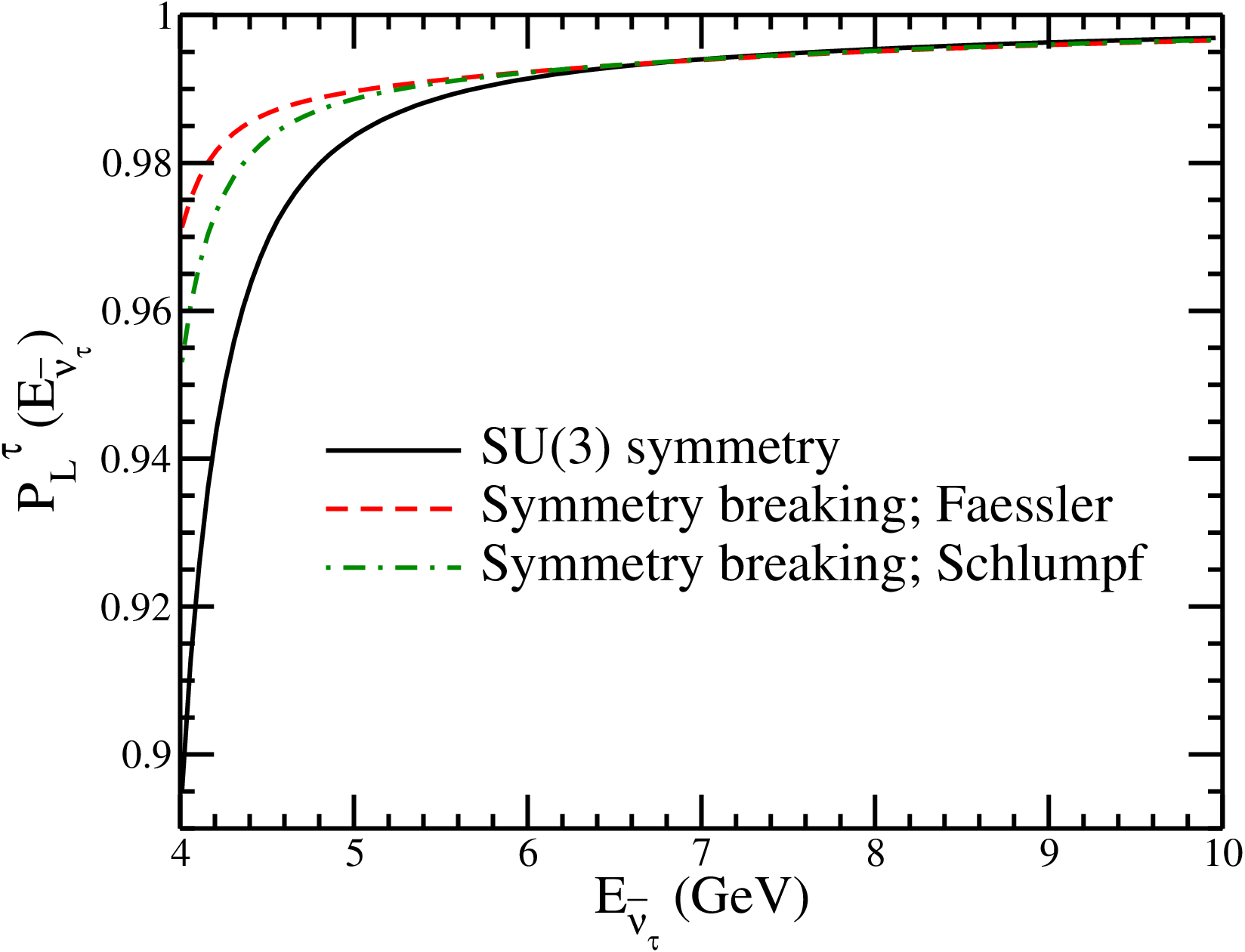}
\includegraphics[height=7cm,width=5.2cm]{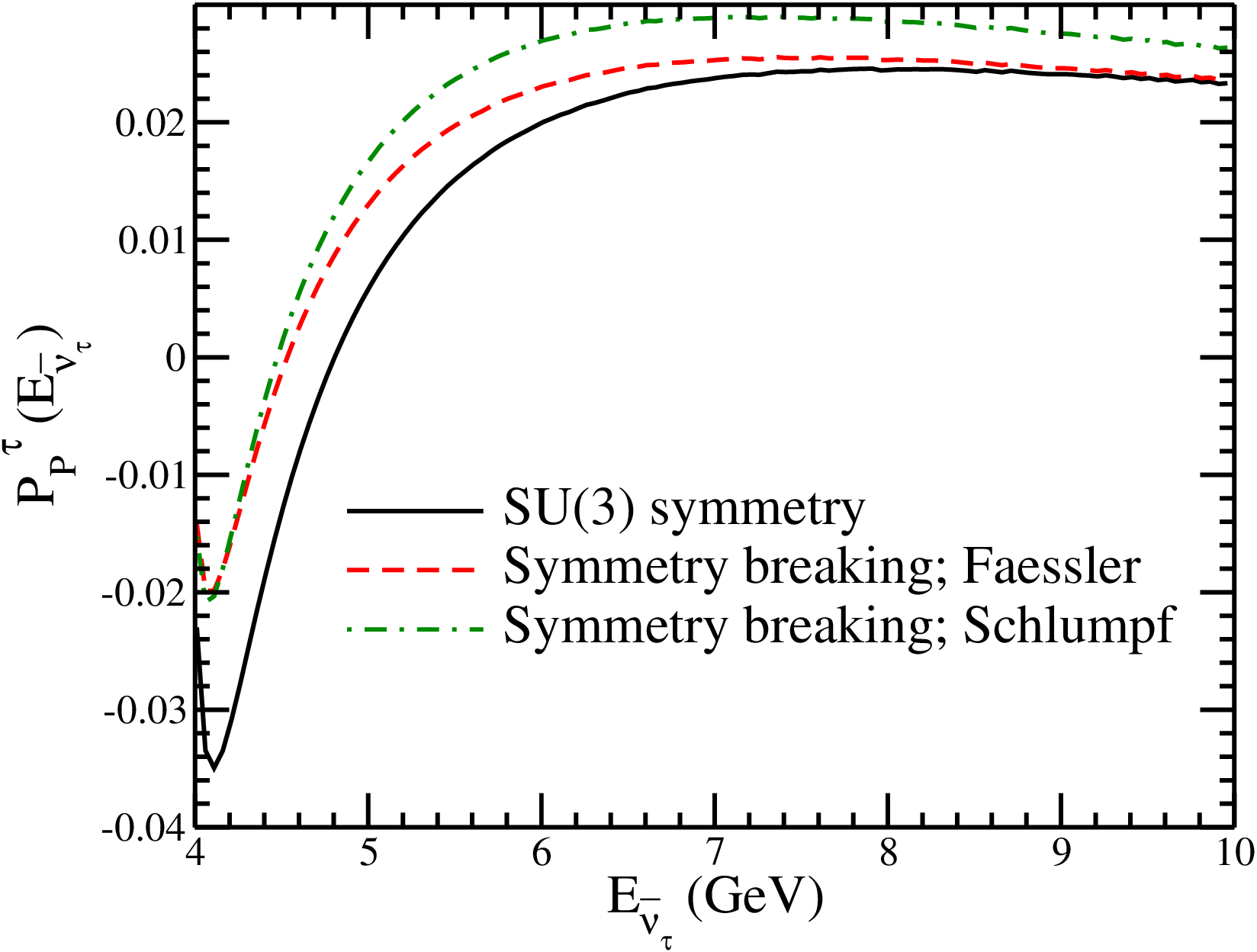}
\caption{$\sigma$~(left panel), $\overline{P}_{L} (E_{\bar{\nu}_{\tau}})$~(middle panel), and $\overline{P}_{P} 
(E_{\bar{\nu}_{\tau}})$~(right panel) vs $E_{\bar{\nu}_{\tau}}$ for $\bar{\nu}_{\tau} + p \rightarrow \tau^{+} + \Lambda$ 
process. The calculations have been performed using the SU(3) symmetry~(solid line), the SU(3) symmetry breaking effects 
parameterized by Faessler {\it et al.}~\cite{Faessler:2008ix}~(dashed line) and by Schlumpf~\cite{Schlumpf}~(dashed-dotted 
line), with electric and magnetic Sachs form factors parameterized by Bradford {\it et al.}~\cite{Bradford:2006yz} and for 
the axial form factor, $M_{A} = 1.026$~GeV is used.}\label{sigma_VFF_lambda_SU3}
\end{figure}

\begin{figure}
\centering
\includegraphics[height=7cm,width=5.2cm]{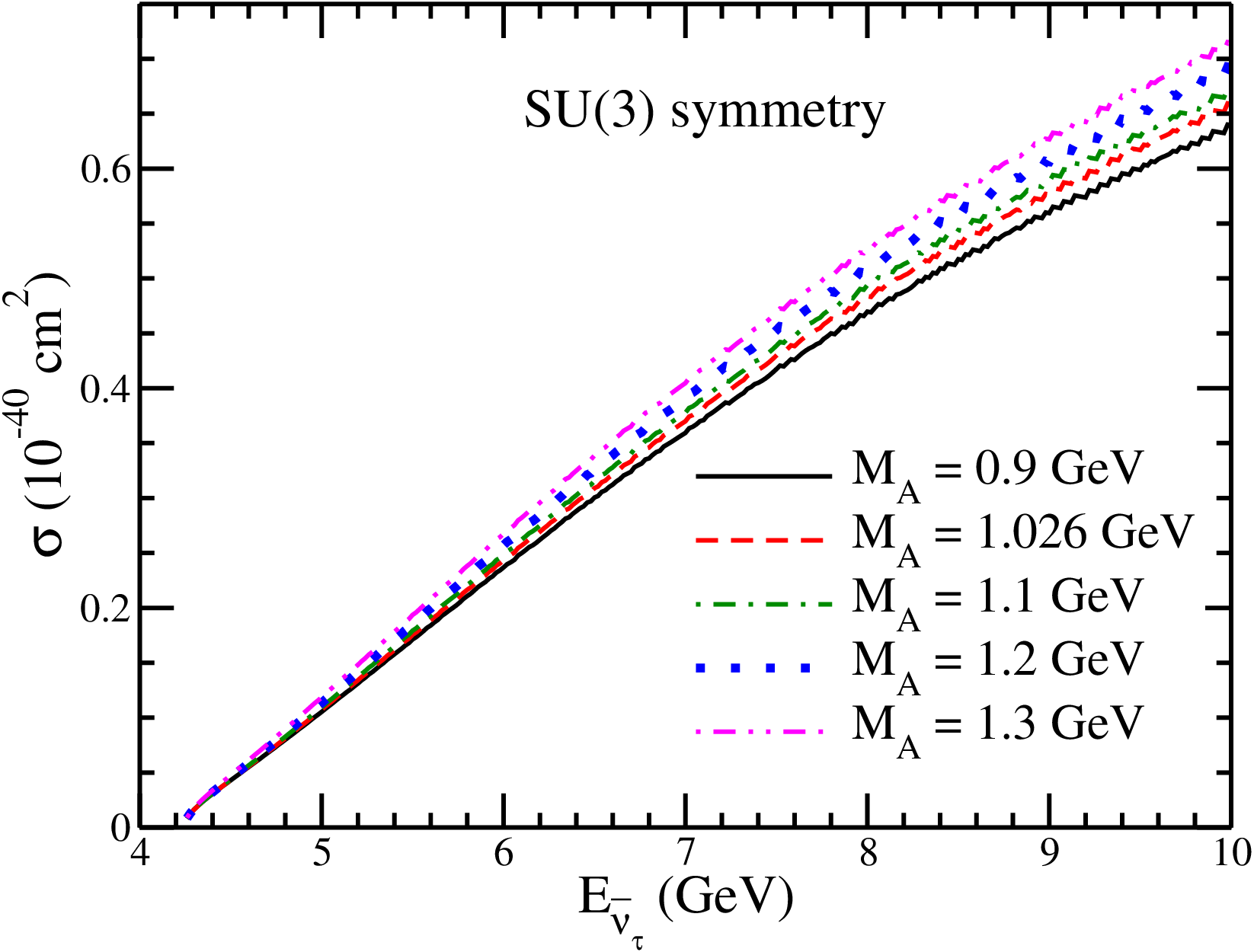}
\includegraphics[height=7cm,width=5.2cm]{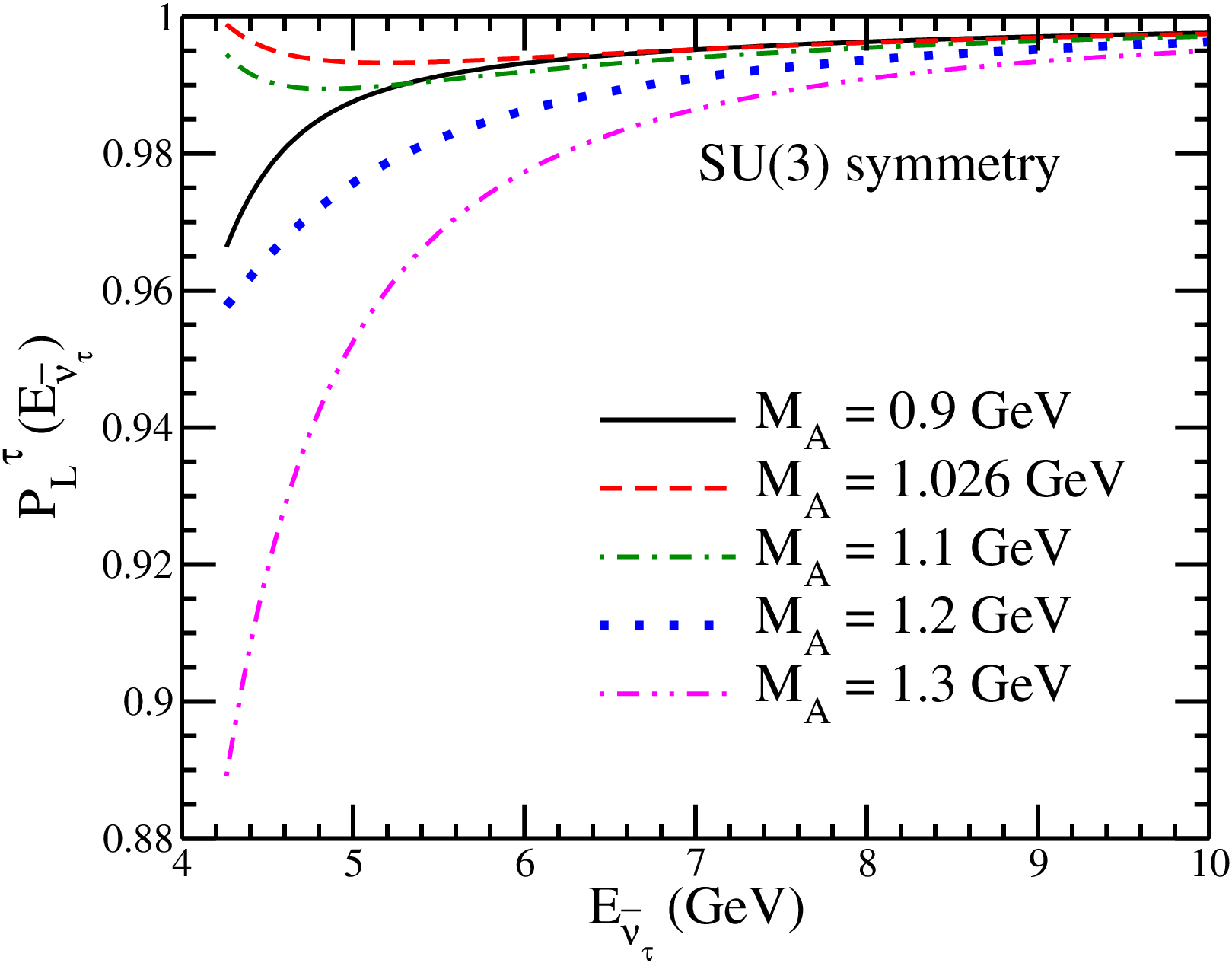}
\includegraphics[height=7cm,width=5.2cm]{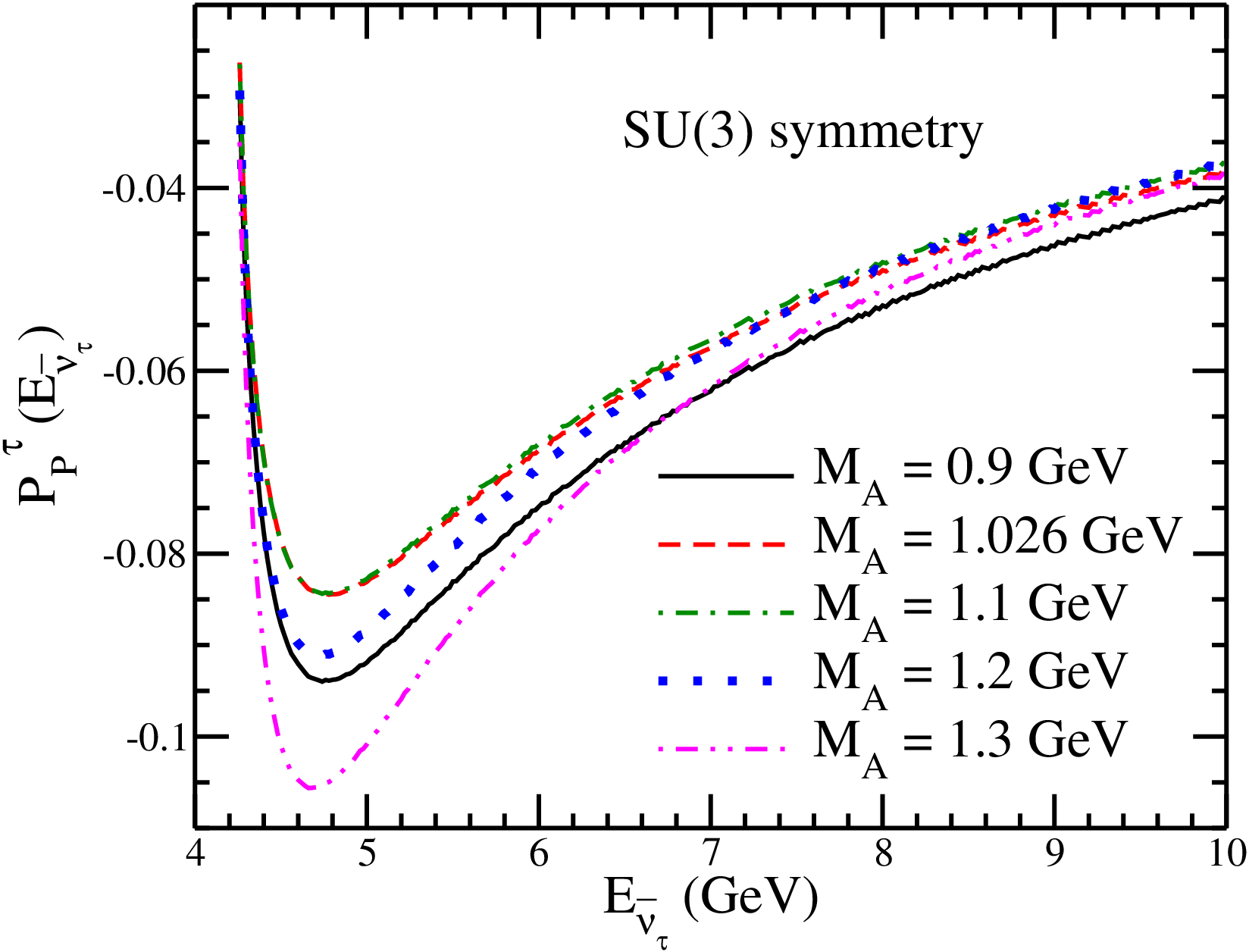}
\caption{$\sigma$~(left panel), $\overline{P}_{L} (E_{\bar{\nu}_{\tau}})$~(middle panel), and $\overline{P}_{P} 
(E_{\bar{\nu}_{\tau}})$~(right panel) vs $E_{\bar{\nu}_{\tau}}$ for $\bar{\nu}_{\tau} + n \rightarrow \tau^{+} + \Sigma^-$ 
process. The calculations have been performed using the SU(3) symmetry with the electric and magnetic Sachs form factors 
parameterized by Bradford {\it et al.}~\cite{Bradford:2006yz} and for the axial form factor, the different values of 
$M_{A}$ have been used {\it viz.} $M_{A} =$ 0.9 GeV~(solid line), 1.026 GeV~(dashed line), 1.1 GeV~(dashed-dotted line), 
1.2 GeV~(dotted line) and 1.3 GeV~(double-dotted-dashed line).}\label{sigma_MA_sigma}
\end{figure}

\begin{figure}
\centering
\includegraphics[height=7cm,width=5.2cm]{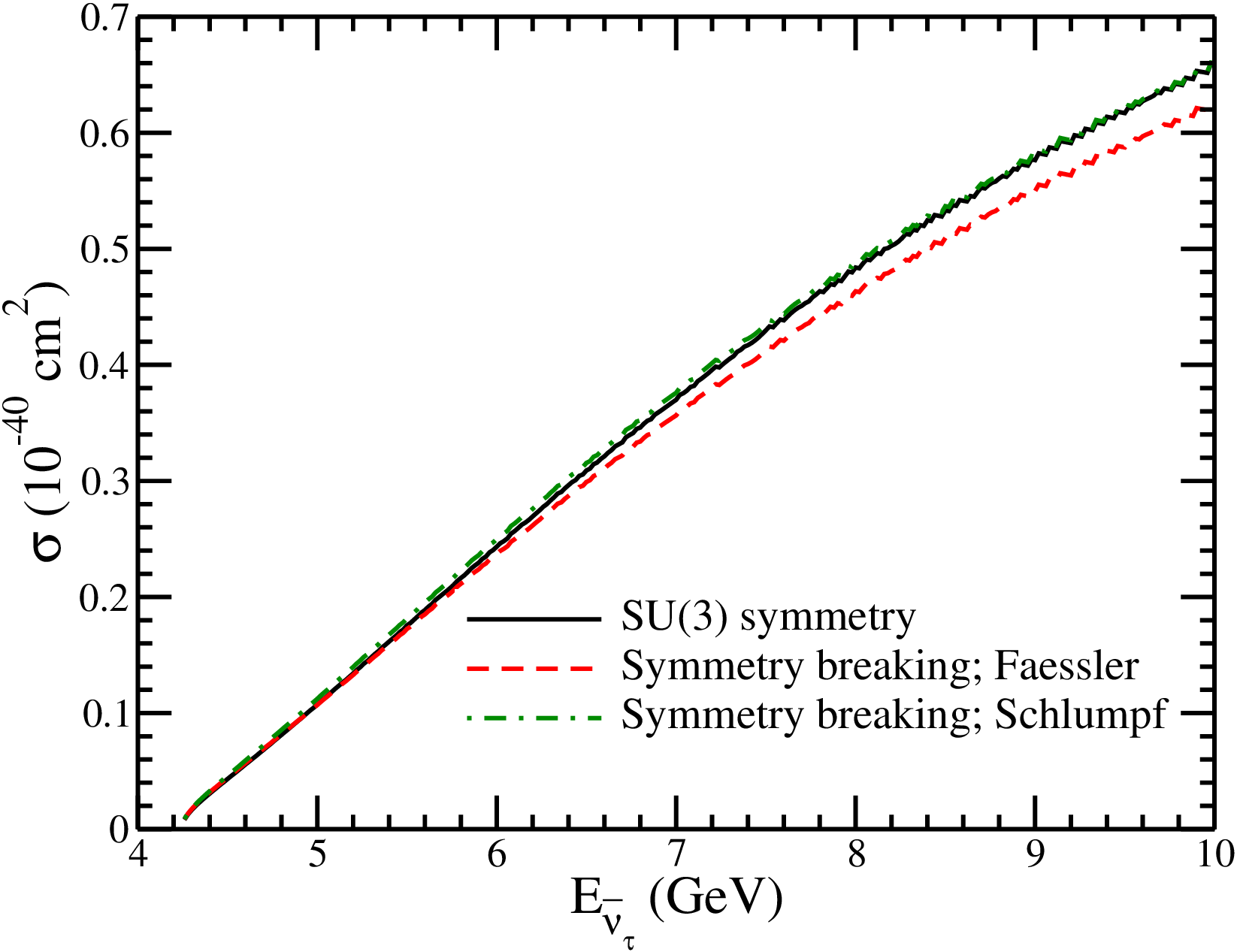}
\includegraphics[height=7cm,width=5.2cm]{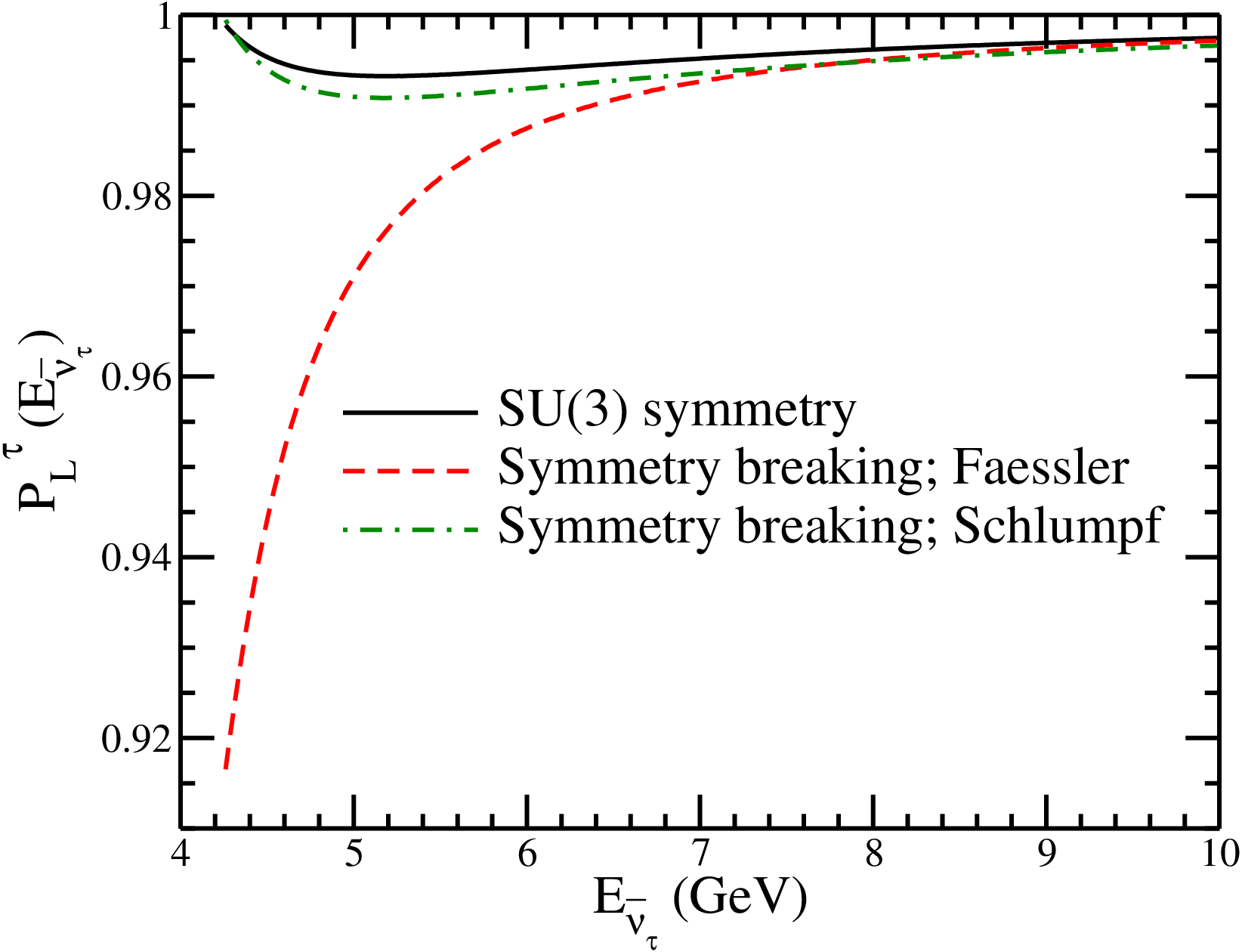}
\includegraphics[height=7cm,width=5.2cm]{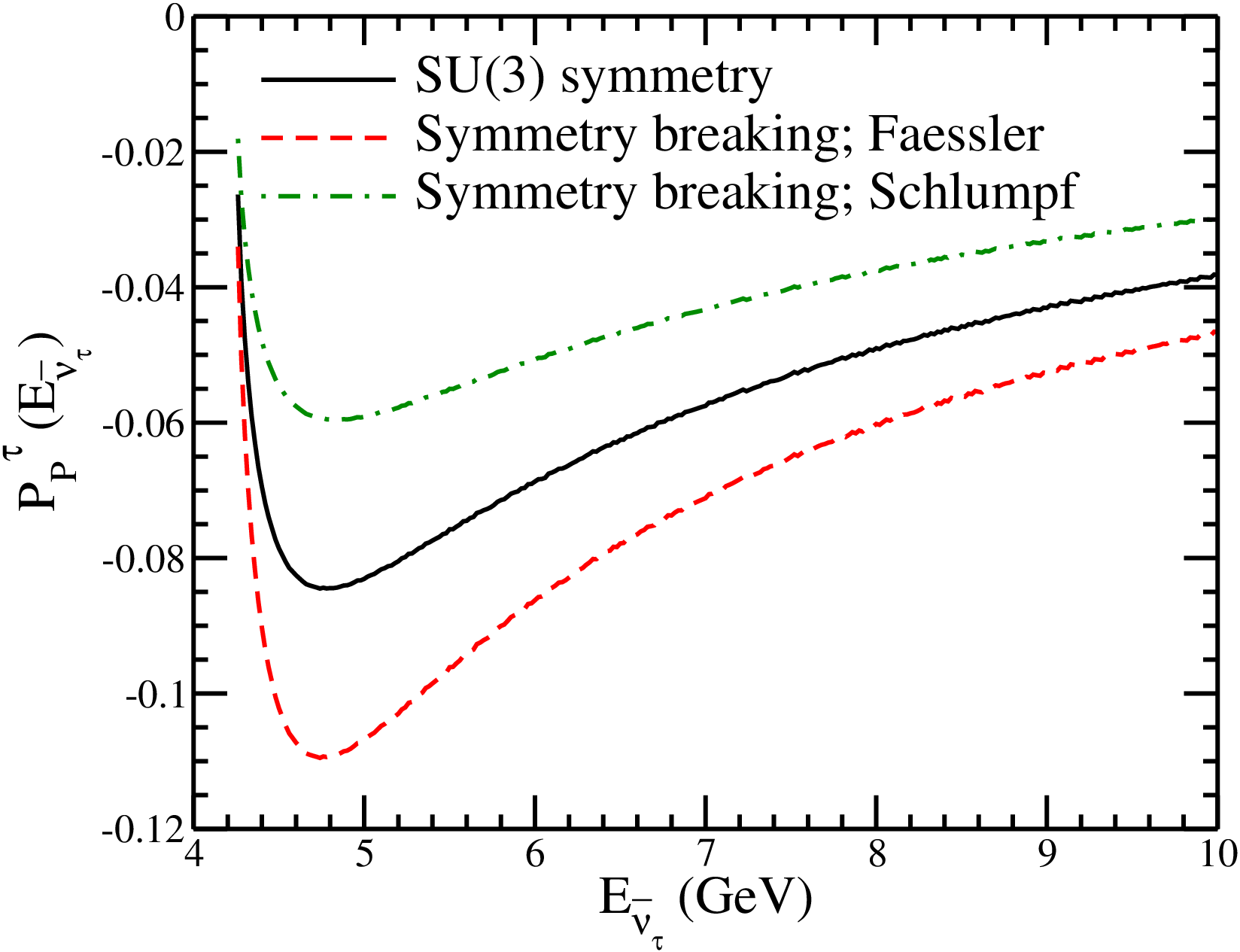}
\caption{$\sigma$~(left panel), $\overline{P}_{L} (E_{\bar{\nu}_{\tau}})$~(middle panel), and $\overline{P}_{P} 
(E_{\bar{\nu}_{\tau}})$~(right panel) vs $E_{\bar{\nu}_{\tau}}$ for $\bar{\nu}_{\tau} + n \rightarrow \tau^{+} + \Sigma^{-}$ 
process. The calculations have been performed using the SU(3) symmetry~(solid line), the SU(3) symmetry breaking effects 
parameterized by Faessler {\it et al.}~\cite{Faessler:2008ix}~(dashed line) and by Schlumpf~\cite{Schlumpf}~(dashed-dotted 
line), with electric and magnetic Sachs form factors parameterized by Bradford {\it et al.}~\cite{Bradford:2006yz} and for 
the axial form factor, $M_{A} = 1.026$~GeV is used.}\label{sigma_VFF_sigma_SU3}
\end{figure}

\subsection{Total scattering cross section and average polarizations}\label{total_cross_section-2}
To study the dependence of the total scattering cross section $\sigma(E_{\bar{\nu}_\tau})$ and the average polarizations 
$\overline{P}_{L,P}(E_{\bar{\nu}_{\tau}})$ on $E_{\bar{\nu}_\tau}$, we have integrated $d\sigma/dQ^2$ and $P_{L,P} 
(Q^2)$ over $Q^2$, and obtained the expressions for $\sigma (E_{\bar{\nu}_\tau})$ and $\overline{P}_{L,P} 
(E_{\bar{\nu}_{\tau}})$ {\it i.e.}:
\begin{eqnarray}\label{total_sig}
\sigma (E_{\bar{\nu}_\tau}) &=& \int_{Q^2_{min}}^{Q^2_{max}} \frac{d\sigma}{dQ^2} dQ^2,\\
\label{average_Plpt}
 \overline{P}_{L,P} (E_{\bar{\nu}_\tau}) &=& \frac{\int_{Q^2_{min}}^{Q^2_{max}} P_{L,P} (Q^2) \frac{d\sigma}{dQ^2} dQ^2}
 {\int_{Q^2_{min}}^{Q^2_{max}} \frac{d\sigma}{dQ^2} dQ^2}.
\end{eqnarray}
In this section, we present the results for the total cross section~(Eq.~(\ref{total_sig})) and average 
polarizations~(Eq.~\ref{average_Plpt}) of the tau lepton produced in $\bar{\nu}_{\tau} + N \longrightarrow \tau^{+} + Y$ 
reaction, separately for $\Lambda$ and $\Sigma^{-}$ productions, in Sections~\ref{cross:lam} and \ref{cross:sig}, respectively.

\subsubsection{$\Lambda$ production}\label{cross:lam}
In Fig.~\ref{sigma_MA_lambda}, the results are presented for $\sigma$ as well as for $\overline{P}_{L} (E_{\bar{\nu}_{\tau}})$ 
and $\overline{P}_{P} (E_{\bar{\nu}_{\tau}})$ obtained using the different values of $M_{A}$ {\it viz.} $M_{A} =$ 0.9, 1.026, 
1.1, 1.2 and 1.3~GeV for the process $\bar{\nu}_{\tau} + p \longrightarrow \tau^{+} + \Lambda$. As expected, $\sigma$ 
increases with the increase in antineutrino energy as well as it increases in magnitude with higher values of $M_{A}$. For 
example, at $E_{\bar{\nu}_{\tau}}=10$~GeV, $\sigma$ increases by about 25\%, when the value of $M_{A}$ is increased from its 
world average value~($M_{A}=1.026$~GeV) by 30\% and a decrease in the value of $M_{A}$ by 10\% from the world average value, 
decreases the cross section by $\sim 7\%$. At low antineutrino energies $E_{\bar{\nu}_{\tau}}=4$~GeV, $\overline{P}_{L} 
(E_{\bar{\nu}_{\tau}})$ increases with the increase in $M_{A}$, which is almost 40\% when $M_{A}$ is varied from 0.9 to 1.3 
GeV. However, with the increase in $E_{\bar{\nu}_{\tau}}$, this variation in $\overline{P}_{L} (E_{\bar{\nu}_{\tau}})$ 
decreases and becomes almost negligible at $E_{\bar{\nu}_{\tau}}=10$~GeV. We observe some dependence of $M_{A}$ on 
$\overline{P}_{P} (E_{\bar{\nu}_{\tau}})$ in the entire range of $E_{\bar{\nu}_{\tau}}$. Note that in the case of $\Lambda$ 
production, $\sigma$, $\overline{P}_{L} (E_{\bar{\nu}_{\tau}})$ and $\overline{P}_{P} (E_{\bar{\nu}_{\tau}})$ are almost 
insensitive to the different parameterizations of the Sachs' form factors, and are not shown here explicitly.
 
In Fig.~\ref{sigma_FP_lambda}, we show the dependence of $\sigma$ and $\overline{P}_{L,P} (E_{\bar{\nu}_{\tau}})$ on 
the pseudoscalar form factor, when the parameterizations by Marshak {\it et al.}~\cite{Marshak} and 
Nambu~\cite{Nambu:1960xd} are used in the numerical calculations. It may be observed from the figure that $\sigma$ as well 
as the average polarizations show some dependence on the pseudoscalar form factor.

In Fig.~\ref{sigma_VFF_lambda_SU3}, we present the results for $\sigma$, $\overline{P}_L (E_{\bar{\nu}_{\tau}})$ and 
$\overline{P}_P (E_{\bar{\nu}_{\tau}})$ with SU(3) symmetry as well as when the SU(3) symmetry breaking effects are taken 
into account following the prescriptions of Faessler 
{\it et al.}~\cite{Faessler:2008ix} and Schlumpf~\cite{Schlumpf}. We find that there is not much effect of SU(3) symmetry 
breaking on $\sigma$, while there is some effect on $\overline{P}_L (E_{\bar{\nu}_{\tau}})$ whereas in the case of 
$\overline{P}_P (E_{\bar{\nu}_{\tau}})$, the effect is found to be significant.

\subsubsection{$\Sigma^{-}$ production}\label{cross:sig}
The dependence of the total cross section and the average polarizations on $M_{A}$ are shown in Fig.~\ref{sigma_MA_sigma}, 
where we present the results for $\sigma$ and $\overline{P}_{L,P} (E_{\bar{\nu}_{\tau}})$ for the process $\bar{\nu}_{\tau} 
+ n \longrightarrow \tau^{+} + \Sigma^{-}$ by varying $M_{A}$ in the range, $M_{A}=$~0.9 to 1.3~GeV. The results 
are qualitatively similar to the results obtained in Fig.~\ref{sigma_MA_lambda} for the $\bar{\nu}_{\tau}$ induced $\Lambda$ 
production. $\sigma$ increases with increase in the value of $M_{A}$ and shows considerable variation at higher antineutrino 
energies. The average polarizations are quite sensitive to the variation in $M_{A}$, especially at low $E_{\bar{\nu}_{\tau}}$.

In Fig.~\ref{sigma_VFF_sigma_SU3}, we show the results for $\sigma$, $\overline{P}_L (E_{\bar{\nu}_{\tau}})$ and 
$\overline{P}_P (E_{\bar{\nu}_{\tau}})$ with SU(3) symmetry as well as when the SU(3) symmetry breaking effects are taken 
into account. We find the effect of SU(3) symmetry breaking to be quite small on $\sigma$. In the case of $\overline{P}_L 
(E_{\bar{\nu}_{\tau}})$, the effect of SU(3) symmetry breaking is more when Faessler {\it et al.}~\cite{Faessler:2008ix} 
prescription is used. While in the case of $\overline{P}_P (E_{\bar{\nu}_{\tau}})$, the symmetry breaking effects using the 
model of Faessler {\it et al.}~\cite{Faessler:2008ix} increases the magnitude of $\overline{P}_P (E_{\bar{\nu}_{\tau}})$ while 
the model of Schlumpf~\cite{Schlumpf} decreases the magnitude of $\overline{P}_P (E_{\bar{\nu}_{\tau}})$ by about 25\% in the 
entire range of $E_{\bar{\nu}_{\tau}}$.

\begin{figure}
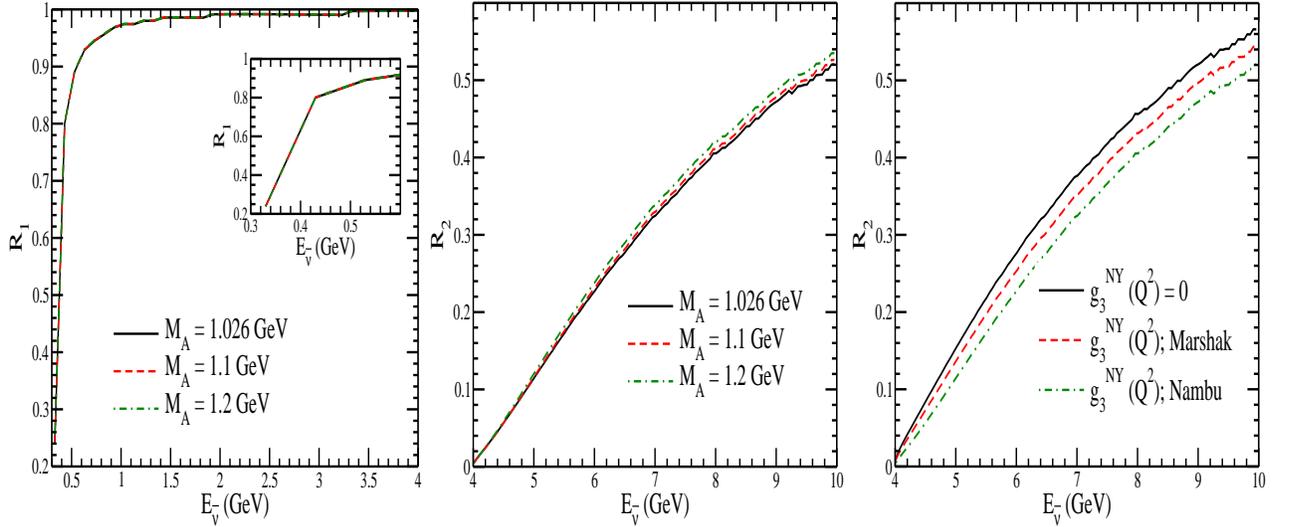

\centering
\includegraphics[height=7cm,width=5.5cm]{ratio_emu_MA.eps}
\includegraphics[height=7cm,width=5.5cm]{ratio_MA.eps}
\includegraphics[height=7cm,width=5.5cm]{ratio_Fp.eps}
\caption{$R_{1}$ as a function of $E_{\bar{\nu}}$~(left panel) with $M_{A} = 1.026$~GeV~(solid line), $M_{A} = 
1.1$~GeV~(dashed line) and $M_{A} = 1.2$~GeV~(dashed-dotted line). $R_{2}$ as a function of $E_{\bar{\nu}}$~(middle panel)
with $M_{A} = 1.026$~GeV~(solid line), $M_{A} = 1.1$~GeV~(dashed line) and $M_{A} = 1.2$~GeV~(dashed-dotted line). $R_{2}$ 
as a function of $E_{\bar{\nu}}$~(right panel) with $g_{3}^{NY} (Q^2)=0$~(solid line), the parameterization of $g_{3}^{NY} 
(Q^2)$ by Marshak {\it et al.}~\cite{Marshak}~(dashed line) and  by Nambu~\cite{Nambu:1960xd}~(dashed-dotted line).}
\label{ratio_SU3}
\end{figure}

\subsection{Implications of lepton flavor universality in neutrino scattering}\label{LFU_emu}
The lepton flavor universality in the $e-\mu$ sector was proposed long time back in 1947 by 
Pontecorvo~\cite{Pontecorvo:1947vp} and has been established phenomenologically by studying the weak processes of $\mu$ 
decay, $\mu^{-}$ capture and $e^-$ capture from nucleons and nuclei~\cite{Athar:2020kqn}. After the discovery of $\tau$ 
lepton and the analyses of its leptonic and semileptonic decays, the principle of lepton flavor universality was extended to 
the $e-\mu-\tau$ sector. As mentioned in the introduction in Section~\ref{intro}, recently there has been considerable work 
in studying the implications of lepton flavor universality in weak interactions in the $e-\mu-\tau$ sector, which have 
focused mainly on the decay processes~\cite{Zhang:2019kky, Yang:2018qdx, Fleischer:2019wlx, Golz:2021yqz, BESIII:2021ynj, 
BESIII:2018ccy, Mezzadri:2018axu, Bifani:2018zmi, Albrecht:2021tul, LHCb:2021trn, deSimone:2020kwi, Belle:2019rba, 
Celani:2021hni}. While there has been very little work on the study of lepton flavor universality in the scattering processes 
induced by $\nu_{e}$, $\nu_{\mu}$ and ${\nu}_{\tau}$. Only two groups have investigated the implications of lepton flavor 
universality in the $e-\mu$ sector by studying the quasielastic scattering induced by $\nu_{e}~(\bar{\nu}_{e})$ and 
$\nu_{\mu}~(\bar{\nu}_{\mu})$ from the free nucleons~\cite{Day:2012gb} and nuclei~\cite{Akbar:2015yda}. They have reported 
the results comparing the cross sections of quasielastic scattering of $\nu_{e} (\bar{\nu}_{e})$ and $\nu_{\mu} 
(\bar{\nu}_{\mu})$ with free nucleons~\cite{Day:2012gb} and with nuclear targets~\cite{Akbar:2015yda} induced by the 
$\Delta S=0$ weak charged current reactions. In Ref.~\cite{Day:2012gb}, Day and McFarland have assumed the LFU and studied 
the differences in the electron and muon production cross section arising due to the lepton mass effect and other effects 
that depend upon the lepton mass like the radiative corrections, the pseudoscalar form factor as well as the form factors 
associated with the second class currents. In Ref.~\cite{Akbar:2015yda}, Akbar {\it et al.} have studied these differences 
in the neutrino-nucleus cross sections including the nuclear medium effects, which are important in the intermediate energy 
region where most of the present neutrino experiments are being done. 

In this section, we study the implications of LFU in the $\Delta S=1$ sector of antineutrino scattering, and compare the 
total cross sections for $e$, $\mu$ and $\tau$ productions in the quasielastic scattering of $\bar{\nu}_{e}$, 
$\bar{\nu}_{\mu}$ and $\bar{\nu}_{\tau}$ from the nucleons induced by the weak charged currents. Specifically, we study the 
ratios of the total cross sections $R_{1}$ and $R_{2}$, defined as: 
\begin{eqnarray}\label{r1}
 R_{1} &=& \frac{\sigma(\bar{\nu}_{\mu} + p \longrightarrow \mu^{+} + \Lambda)}{\sigma(\bar{\nu}_{e} + p \longrightarrow e^{+} 
 + \Lambda)},\\
 \label{r2}
 R_{2} &=& \frac{2\sigma(\bar{\nu}_{\tau} + p \longrightarrow \tau^{+} + \Lambda)}{\sigma(\bar{\nu}_{\mu} + p \longrightarrow 
 \mu^{+} + \Lambda) + \sigma(\bar{\nu}_{e} + p \longrightarrow e^{+} + \Lambda)},
\end{eqnarray}
as a function of antineutrino energy $E_{\bar{\nu}}$ and investigate the effect of axial dipole mass $M_{A}$ and the 
pseudoscalar form factor $g_{3}^{NY} (Q^2)$ assuming the SU(3) symmetry. The numerical results are calculated by taking 
$M_{A}=1.026$~GeV and the parameterization of $g_{3}^{NY} (Q^2)$ by Nambu, unless stated otherwise, and presented in 
Fig.~\ref{ratio_SU3}. We have also studied the effect of changing the vector form factors as well as the effect of SU(3) 
symmetry breaking. These effects are found to be quantitatively very small on $R_{1} (E_{\bar{\nu}})$ and $R_{2} 
(E_{\bar{\nu}})$, and are not presented here.\\

We see from Fig.~\ref{ratio_SU3} that
\begin{itemize}
 \item [(i)] the ratios $R_{1} (E_{\bar{\nu}})$~(Eq.~(\ref{r1})) and $R_{2} (E_{\bar{\nu}})$~(Eq.~(\ref{r2})) have very 
 little dependence on the choice of $M_{A}$, which is almost negligible in the case of $R_{1} (E_{\bar{\nu}})$. It may be 
 noticed that in the kinematic region of $\tau$ production {\it i.e.}, $E_{\bar{\nu}} \simeq 4$~GeV, the ratio $R_{1} 
 (E_{\bar{\nu}})$ is almost unity. While in the case of $R_{2} (E_{\bar{\nu}})$, the ratio is highly suppressed due to the 
 threshold effects and becomes more than 0.5, only for $E_{\bar{\nu}}> 10$~GeV. Thus, any deviation of $R_{1} 
 (E_{\bar{\nu}})$ and $R_{2} (E_{\bar{\nu}})$ from the values shown in Fig.~\ref{ratio_SU3}, would be a possible signal for 
 the violation of LFU.
 
 \item [(ii)] the ratio $R_{2} (E_{\bar{\nu}})$ has some dependence on the choice of the pseudoscalar form factor $g_{3}^{NY} 
 (Q^2)$. If we take $g_{3}^{NY} (Q^2)=0$ or the parameterization of $g_{3}^{NY} (Q^2)$ given by Marshak {\it et 
 al.}~\cite{Marshak}, the value of $R_{2} (E_{\bar{\nu}})$ increases as compared to the value obtained by using the 
 parameterization of $g_{3}^{NY} (Q^2)$ given by Nambu~\cite{Nambu:1960xd}. For example, at $E_{\bar{\nu}} =10$~GeV, when we 
 compare the results obtained with $g_{3}^{NY} (Q^2)$ using the parameterization of Marshak {\it et al.}~\cite{Marshak}, the 
 value of $R_{2} (E_{\bar{\nu}})$ increases by 5\%, which becomes 18\% at $E_{\bar{\nu}} =5$~GeV, from the results using 
 $g_{3}^{NY} (Q^2)$ from the results obtained using $g_{3}^{NY} (Q^2)$ parameterized by Marshak {\it et al.}~\cite{Marshak}. 
 Therefore, an experimental determination of $R_{2} (E_{\bar{\nu}})$ with a precision of 20\% or higher would be able to show 
 any evidence of the violation of LFU in the $e-\mu-\tau$ sector.
 
\end{itemize}

\section{Summary and conclusions}\label{summary}
In this work, we have presented the results for the $|\Delta S|=1$ hyperon production in the quasielastic 
${\bar\nu}_\tau$-nucleon scattering and obtained the differential~($\frac{d\sigma}{dQ^2}$) and total~($\sigma$) scattering 
cross sections as well as the longitudinal~($P_L$) and perpendicular~($P_P$) components of the polarized $\tau^{+}$ lepton 
produced in these reactions. We studied theoretical uncertainties arising due to the use of different vector, axial vector and 
pseudoscalar form factors as well as the effect of SU(3) symmetry breaking on these observables.\\

Our observations are the following:
 \begin{itemize}
 \item [(i)] In the case of $\Lambda$ production, the total cross section as well as the average polarizations increases with 
 increase in $M_{A}$ at low antineutrino energies, say at $E_{\bar{\nu}_{\tau}} = 5$~GeV. However, with the increase in 
 $E_{\bar{\nu}_{\tau}}$, $\sigma$ further increases with $M_{A}$ while $\overline{P}_{L,P} (E_{\bar{\nu}_{\tau}})$ saturates 
 with increase in $M_{A}$. Moreover, in the case of $\Sigma^{-}$ production, the results for $\sigma$ as well as 
 $\overline{P}_{L,P} (E_{\bar{\nu}_{\tau}})$ are qualitatively similar to the results obtained for $\Lambda$ production.
 
 \item [(ii)] There is not much effect of SU(3) symmetry breaking on $\sigma$ for both $\Lambda$ and $\Sigma^{-}$ 
 productions. However, we observe some dependence of the symmetry breaking effect on $\overline{P}_{L,P} 
 (E_{\bar{\nu}_{\tau}})$, which are qualitatively different for $\Lambda$ and $\Sigma^{-}$ productions. 
 
 \item [(iii)] In the case of $\Lambda$ production, at low ${\bar\nu}_\tau$ energies, specifically near the threshold energy, 
 the effect of the different parameterizations of vector form factors on $\frac{d\sigma}{dQ^2}$, $P_L(Q^2)$ and $P_P(Q^2)$ 
 distributions is large, which decreases with the increase in $E_{{\bar\nu}_\tau}$. However, in the case of $\Sigma^{-}$ 
 production, $\frac{d\sigma}{dQ^2}$ shows appreciable dependence on the different parameterizations of the vector form factors 
 for all values of $E_{{\bar\nu}_\tau}$, while $P_L(Q^2)$ and $P_P(Q^2)$ distributions are quite sensitive to the choice of 
 the vector form factors. 
 
\item [(iv)] The effect of variation in the axial dipole mass $M_A$ on $\frac{d\sigma}{dQ^2}$, $P_L(Q^2)$ and $P_P(Q^2)$, 
when $\Lambda$ is produced in the final state, is significant at low $E_{{\bar\nu}_\tau}$, which decreases with the increase 
in $E_{{\bar\nu}_\tau}$ for $Q^2$ and $P_L(Q^2)$ distributions but for $P_P(Q^2)$ distribution still remains significantly 
different even at high ${\bar\nu}_\tau$ energies. While there is a large dependence of $M_A$ on $\frac{d\sigma}{dQ^2}$, 
$P_L(Q^2)$ and $P_P(Q^2)$ at all values of $Q^2$ and $E_{\bar{\nu}_\tau}$, when $\Sigma^{-}$ is produced in the final state.

\item [(v)] In the case of $\Lambda$ production, the effect of SU(3) symmetry breaking on $\frac{d\sigma}{dQ^2}$ is very 
small at all values of $E_{\bar{\nu}_{\tau}}$ and $Q^2$ while there is a sizeable effect of SU(3) symmetry breaking on 
$P_L(Q^2)$ and $P_P(Q^2)$. However, in the case of $\Sigma^{-}$ production, the effect of SU(3) symmetry breaking on 
$\frac{d\sigma}{dQ^2}$ is small at low $E_{\bar{\nu}_{\tau}}$, and increases with the increase in energy. Moreover, different 
results for the polarization observables are obtained when the SU(3) symmetry breaking using the prescriptions by 
Schlumpf~\cite{Schlumpf} and Faessler {\it et al.}~\cite{Faessler:2008ix} are taken into account.  

\item [(vi)] We have also tested the lepton flavor universality in the antineutrino induced $\Lambda$ production by
calculating the ratios~(given in Eqs.~(\ref{r1}) and (\ref{r2})) of the total cross sections in the $e-\mu$ and $e-\mu-\tau$ 
sectors. We find that there is no dependence of $M_{A}$ or the SU(3) symmetry breaking on $R_{1}$. However, $R_{2}$ shows 
some dependence on the choice of $M_A$ as well as to the different parameterizations of the pseudoscalar form factor. Thus, 
the experimental observation of $R_{2} (E_{\bar{\nu}})$ with a precision of 20\% or higher would be able to show any evidence 
of the violation of LFU in the $e-\mu-\tau$ sector.

\end{itemize}

\section*{Acknowledgment}   
A. F. and M. S. A. are thankful to the Department of Science and Technology (DST), Government of India for providing 
financial assistance under Grant No. SR/MF/PS-01/2016-AMU.

\end{document}